\documentclass[useAMS,usenatbib,usegraphics,referee]{mn2e}
\usepackage{amssymb}
\usepackage{lscape}
\usepackage{ulem}
\usepackage{txfonts}
\usepackage{natbib}
\usepackage{rotating}
\usepackage{ctable}
\usepackage{url}
\usepackage{fixltx2e} 
\usepackage{longtable}

\def\lea{\mathrel{<\kern-1.0em\lower0.9ex\hbox{$\sim$}}}
\def\gea{\mathrel{>\kern-1.0em\lower0.9ex\hbox{$\sim$}}}

\newcommand{\apj}{ApJ}
\newcommand{\apjl}{ApJL}
\newcommand{\apjs}{ApJS}
\newcommand{\aj}{AJ}
\newcommand{\mnras}{MNRAS}
\newcommand{\nat}{Nature}

\newcommand{\araa}{ARA\&A}
\newcommand{\aap}{A\&A}

\newcommand\aaps{{A\&AS}}

\newcommand\jrasc{J. R. Astron. Soc. Can.}

\title[Newly Identified Star Clusters In M33. III]{Newly Identified Star Clusters in M33. III. Structural Parameters}

\author[San Roman et al.]{I. San Roman$^{1}$\thanks{E-mail: izaskun@astro.ufl.edu},  A. Sarajedini$^{1}$\thanks{E-mail: ata@astro.ufl.edu}, J. A. Holtzman$^{2}$ and  D. R. Garnett$^{3}$\\
$^{1}$Department of Astronomy, University of Florida, 211 Bryant Space Science Center, Gainesville, FL 32611-2055, USA\\
$^{2}$New Mexico State University, Las Cruces, NM, USA\\
$^{3}$801 W. Wheatridge Drive, Tucson AZ 85704, US}

\begin{document}

\pagerange{\pageref{firstpage}--\pageref{lastpage}} \pubyear{2011}

\maketitle

\label{firstpage}

\begin{abstract}
We present the morphological properties of 161 star clusters in M33 using the Advanced Camera For Surveys Wide Field Channel onboard the Hubble Space Telescope using observations with the F606W and F814W filters.  We obtain, for the first time, ellipticities, position angles, and surface brightness profiles for a significant number of clusters. On average, M33 clusters are more flattened than those of the Milky Way and M31, and more similar to clusters in the Small Magellanic Cloud. The ellipticities do not show any correlation with age or mass, suggesting that rotation is not the main cause of elongation in the M33 clusters. The position angles of the clusters show a bimodality with a strong peak perpendicular to the position angle of the galaxy major axis. These results support the notion that tidal forces are the reason for the cluster flattening. We fit King and EFF models to the surface brightness profiles and derive structural parameters including core radii, concentration, half-light radii and central surface brightness for both filters. The surface brightness profiles of a significant number of clusters show irregularities such as bumps and dips. 
Young clusters (Log age $<$ 8) are notably better fitted by models with no radial truncation (EFF models), while older clusters show no significant differences between King or EFF fits. M33 star clusters seem to have smaller sizes, smaller concentrations, and smaller central surface brightness as compared to clusters in the MW, M31, LMC and SMC. Analysis of the structural parameters presents a age-radius relation also detected in other star cluster systems. The overall analysis shows differences in the structural evolution between the M33 cluster system and cluster systems in nearby galaxies. These differences could have been caused by the strong differences in these various environments.

\end{abstract}


\section{Introduction}

The spatial structures and internal stellar kinematics of star clusters (SCs) bear the imprint of their initial conditions and their dynamical evolution. 
Determination of structural parameters of SCs yields important information to understand the processes involved in their formation and evolution. Due to their proximity, galaxies in the Local Group provide us with ideal targets for detailed studies of SC structural parameters. While the SC systems of the Milky Way (MW) and M31 have received close attention, the third spiral galaxy in the Local Group, M33, has been less studied. As pointed out by previous authors \citep{Sarajedini1998,Chandar2006,SanRoman2010}, clusters in M33 populate almost all evolutionary phases, including ages and metallicities not observed in the MW or M31. Therefore, M33 star clusters can provide us with essential information to understand the formation and evolution of their early and late stages. At a distance of 870 Kpc \citep[distance modulus = 24.69;][]{Galletietal2004}, M33 is the only nearby late-type spiral galaxy (Scd) and provides us with an ideal opportunity to study the star cluster system of such a galaxy.

The structure of SCs can be investigated through their surface brightness profiles. These profiles are often characterized by two families of models. The first set, known as King models, are based on tidally limited profiles of isothermal spheres \citep{King1962,King1966}. More extensive King-based models exist that take into consideration non-isotropic systems, rotating systems or even multi-mass models \citep[e.g.][]{Michie1963, Wilson1975,DaCosta1976}. The second family of models, known as EFF (for Elson, Freeman, and Fall) models, were empirically derived to reproduce the surface brightness profiles of young clusters in the Magellanic Clouds \citep{Elson1987,Elson1991}. These profiles are described as power-laws that do not include the pronounced tidal truncation of the King models.

It is well known that King models provide an excellent description of the luminosity profiles and internal kinematics of most MW globular clusters \citep[e.g.][]{Djorgovski1995,Trager1995,McLaughlin2000}. High spatial resolution data are required to study spatial structures in extragalactic clusters. \textit{Hubble Space Telescope} (HST) imaging has been successsfully used to fit a variety of models to the surface brightness profiles of clusters in M31 \citep{Barmby2002,Barmby2007,Barmby2009}, the Magellanic Clouds (MC), and the Fornax dwarf spheroidal galaxy \citep{Mackey2003a,Mackey2003b,Mackey2003c,McLaughlin2005}. These studies suggest that globular clusters (GCs) describe a fundamental plane analogous to the fundamental plane of elliptical galaxies. Morphological studies of younger clusters present more difficulties based on their, generally, lower concentration and mass; it is not well established if open clusters (OCs) in the MW also lie on the same fundamental plane \citep{Bonatto2005,Barmby2009}. Therefore, it is essential to determine the structural parameters  of star clusters in other galaxies to constrain SC initial conditions and evolution.

In the last few years, significant advances have been made to understand the star cluster system of M33 \citep[e.g.][]{SanRoman2009,SanRoman2010,Cockcroft2011}. However, structural studies of this system are very limited. \citet{Larsen2002} present structural parameters derived from King model fits for four GCs in M33. Dynamically and structurally, they appear virtually identical to the MW and M31 GCs, and fit very well into the fundamental plane. \citet{Chandar1999} present core radii of 60 star clusters using linear correlations with the measured full width at half maximum (FWHM) of each cluster. For these reasons, we have undertaken a comprehensive structural study of the largest sample of M33 star clusters to date. Throughout this work, a distance to M33 of 870 Kpc \citep{Galletietal2004} has been adopted. At that distance, 1'' corresponds to 4.22 pc.

This paper presents the results of this study and is organized as follows. Section 2 describes the observations and data reduction. Section 3 presents the analysis of ellipticities and position angles while Section 4 discusses the surface brightness profiles and the profile fittings. The analysis of the structural parameters and comparison with other galaxies are shown in Section 5. Finally, Section 6 presents a summary of the study.

\section[]{Observations and Data Reduction}

The observations and data reduction for the present study are described in a companion paper \citep[hereafter Paper I]{SanRoman2009}, and for a detailed description we refer the reader to that paper. For convenience, we provide an abbreviated summary below.

The observations were obtained with the Advanced Camera for Surveys Wide Field Channel (ACS/WFC) on board the HST.  Twelve \textit{HST}/ACS fields from the GO-10190 program (P.I.: D. Garnett) have been analyzed. Three filters were used for the primary observations (F475W, F606W, F814W) and two for the parallel images (F606W, F814W). The fields are located along the southwest direction covering a radial extension between 0.9 - 6.6 Kpc from the center of the galaxy. Figure 1 in Paper I shows the locations of these fields while Table 1 of that paper presents a summary of the observations.

All of the images were calibrated through the standard pipeline-process. The standard-calibrated ``FLT'' images were corrected for the geometric distortion and then used to obtain the point-spread function (PSF). A detailed description of the PSF construction method is presented in \cite{Sarajedini2006}. This PSF was used to convolve the structural models before the fitting process.

The pipeline-processed drizzled (``DRZ'') images were used to construct the surface brightness photometry. We derived positional offsets between these frames using the \textit{imshift} and \textit{imcombine} tasks in IRAF to allow us to produce one master image per filter per field. To make the calculation of the photometric errors more straightforward, each DRZ image was multiplied by the exposure time before performing the photometry, and then the background sky value was added back to each image.

\section[]{Ellipticities}

Crowding and incompleteness in the inner region of the clusters make star counts impractical. For this reason, we have derived the structural parameters of the clusters from luminosity profiles rather than number density profiles. A precise determination of the cluster centers is crucial to obtain this morphological information of the objects. Inaccurate centers would produce artificial distortions in the radial profiles. To obtain cluster centers and shapes we used the IRAF task ELLIPSE to fit elliptical isophotes to our images in both filters. ELLIPSE is an algorithm designed to fit galaxy profiles that decrease monotonically, so we first filtered the images with an 11 x 11 pixel$^{2}$ median filter to create the required smooth profiles. Spatial smoothing of the images is a common technique when dealing with surface brightness rather than number density of stars, however the required magnitude of the smoothing could degrade the accuracy of the derived centers. At the distance of M33, our objects are partially unresolved so our images do not need a high degree of filtering to obtain an appropriate smooth profiles. This technique has been analyzed and successfully used in previous studies \citep[e.g.][]{Larsen2002,Barmby2002,McLaughlin2008}.  

 Isophotes were fitted between 0''.1 and the largest measurable semi-major axis in both bands. The overall center position of each cluster has been determined as the average of the ELLIPSE output between 0''.5 and 1''.5 along the semi-major axis where the measurements are more reliable. The final averaged position of each cluster differs by less than 0''.1  between the F606W and F814W filters. Because of this small offset in the cluster position between the different filters, we have chosen the cluster coordinates determined from the F606W filter for the following analysis.

\begin{figure*}
\begin{center}
\includegraphics[width=0.49\textwidth]{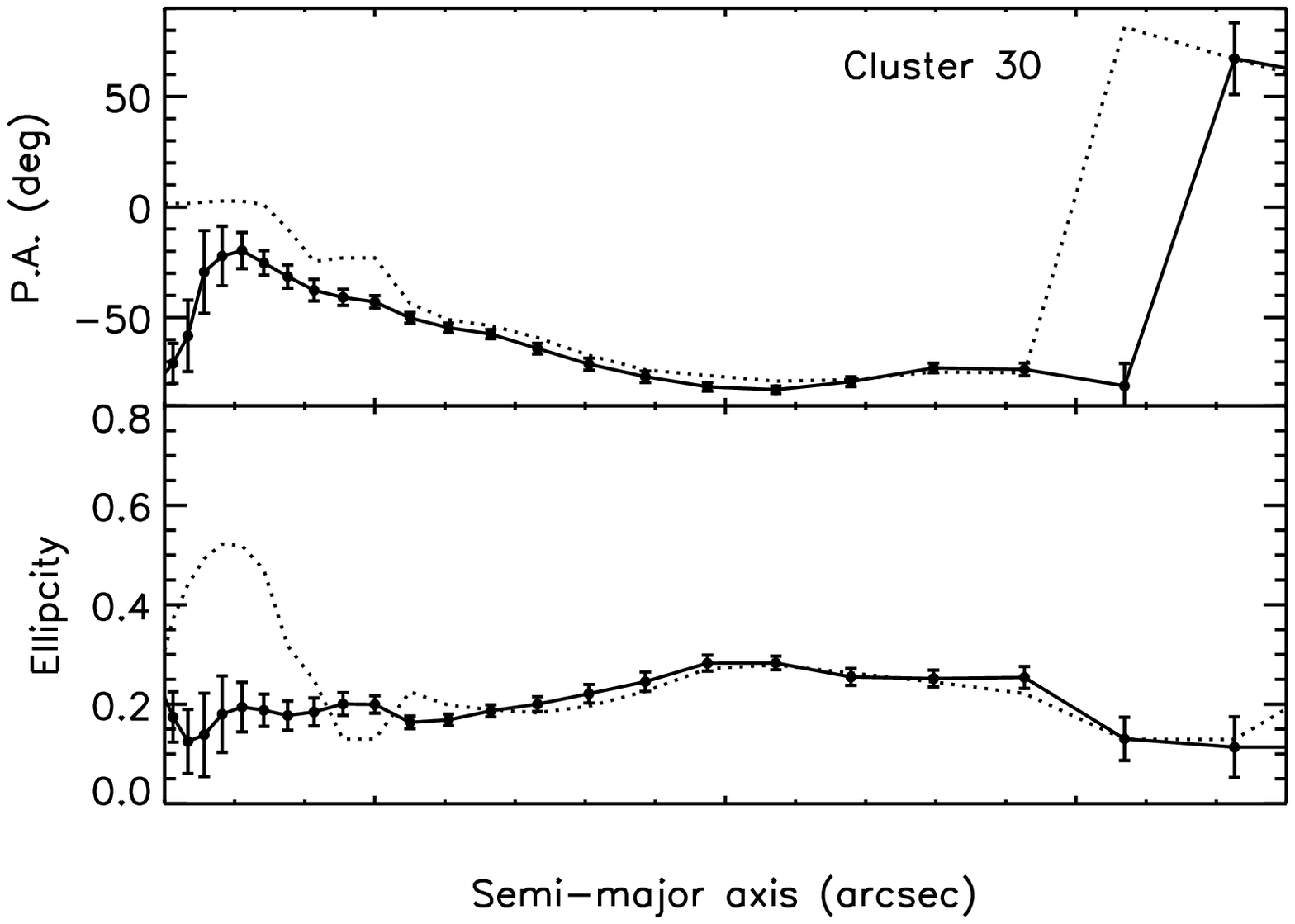} 
\includegraphics[width=0.49\textwidth]{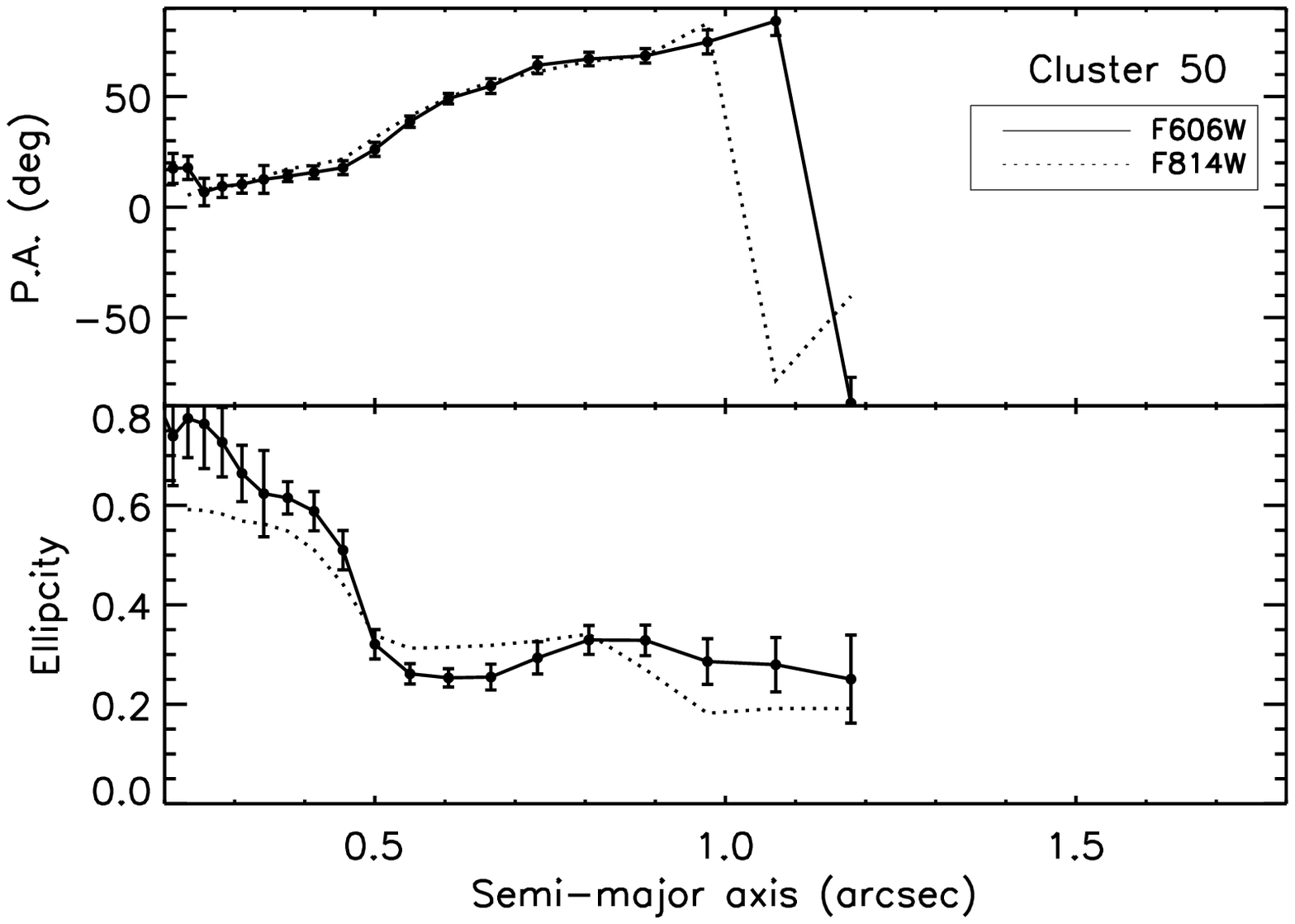} 
\includegraphics[width=0.49\textwidth]{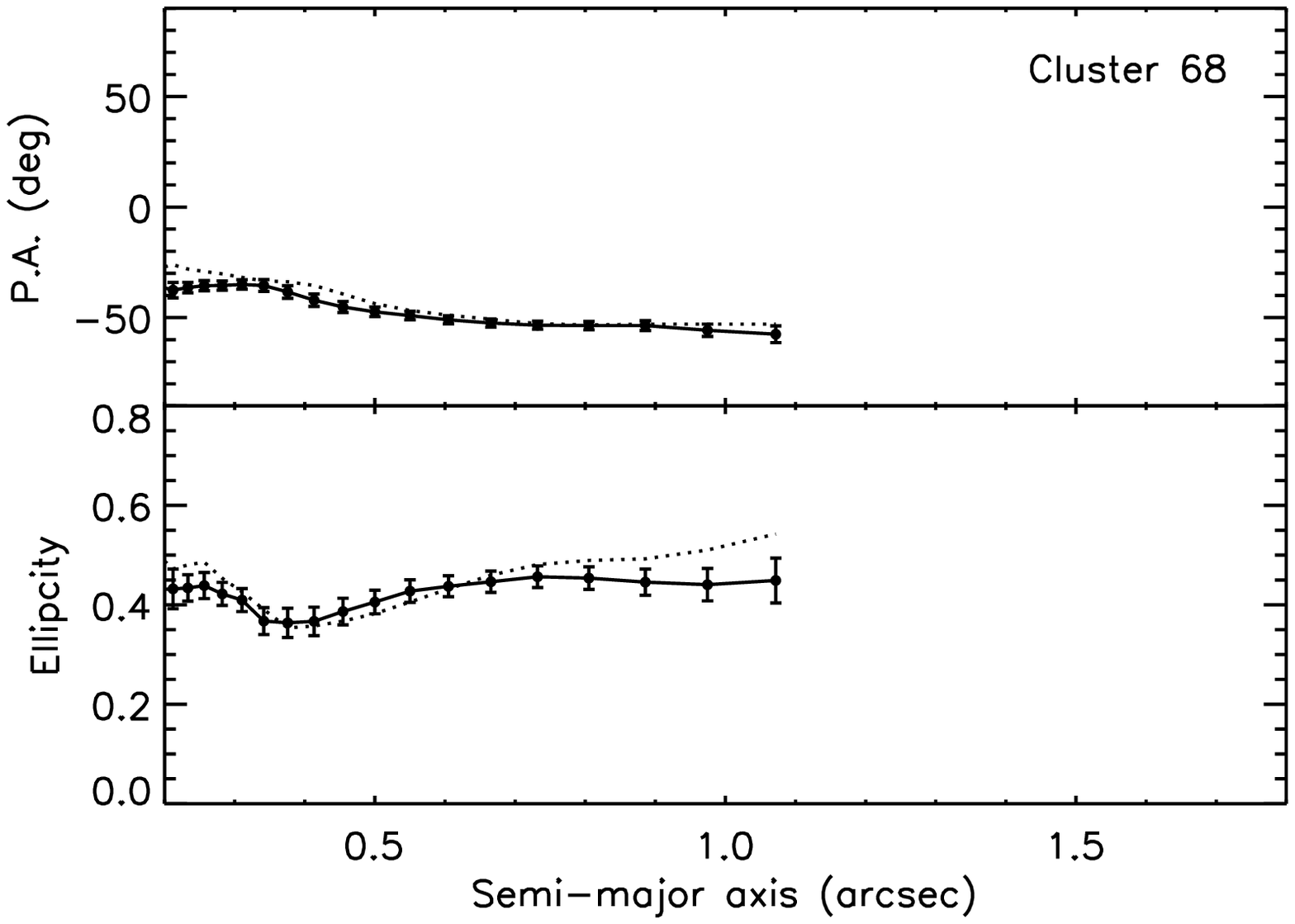} 
\includegraphics[width=0.49\textwidth]{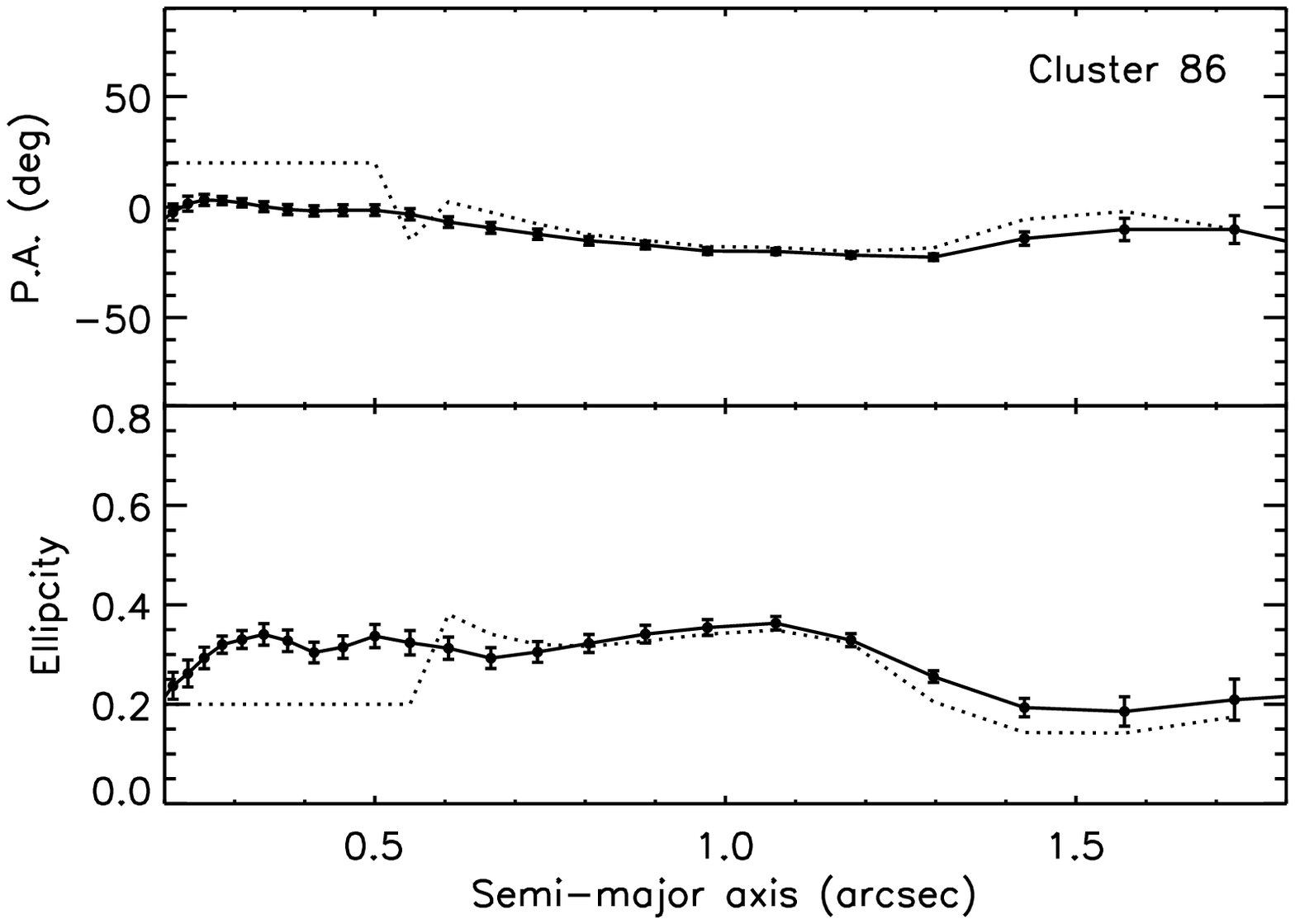}
\includegraphics[width=0.49\textwidth]{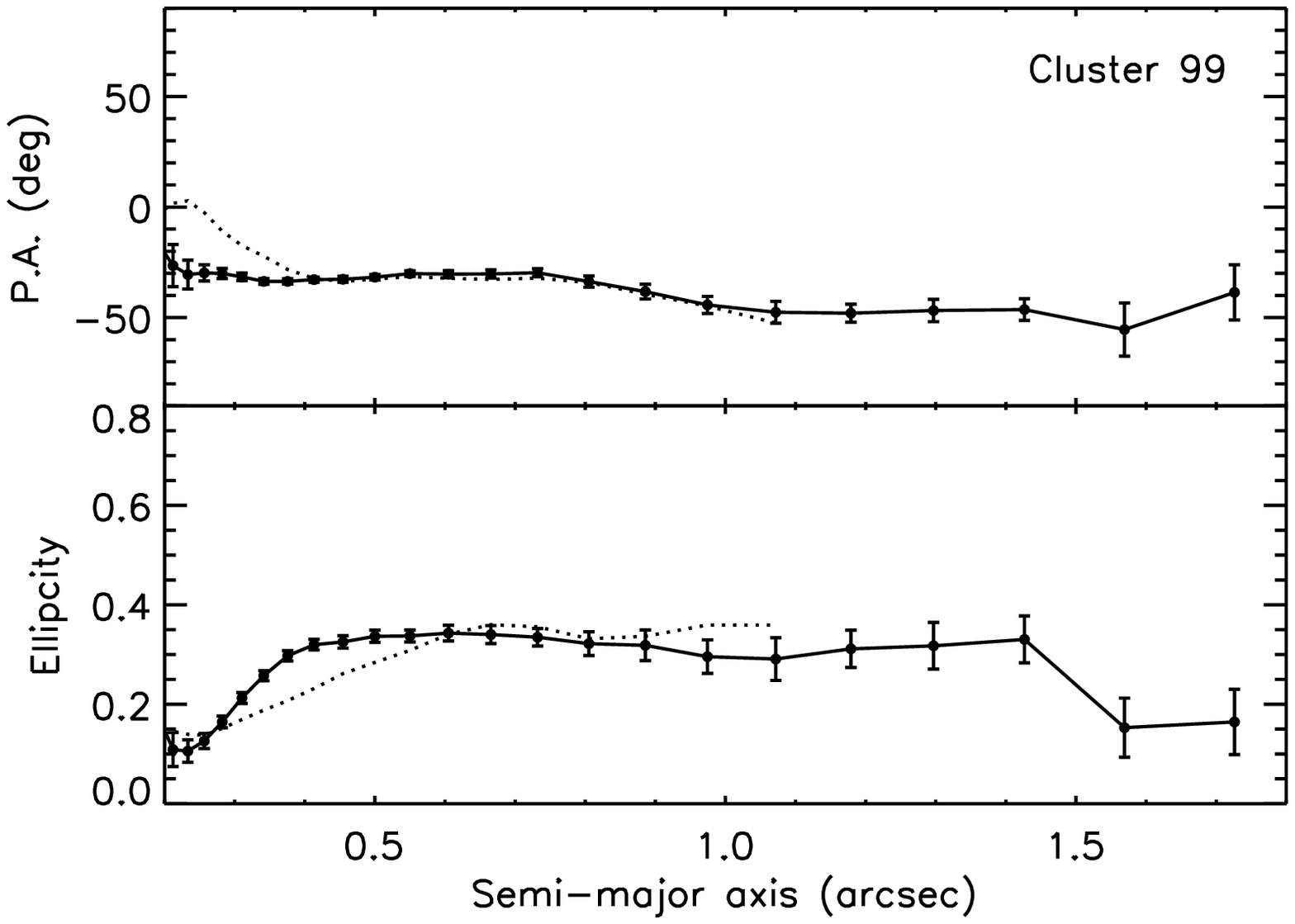} 
\includegraphics[width=0.49\textwidth]{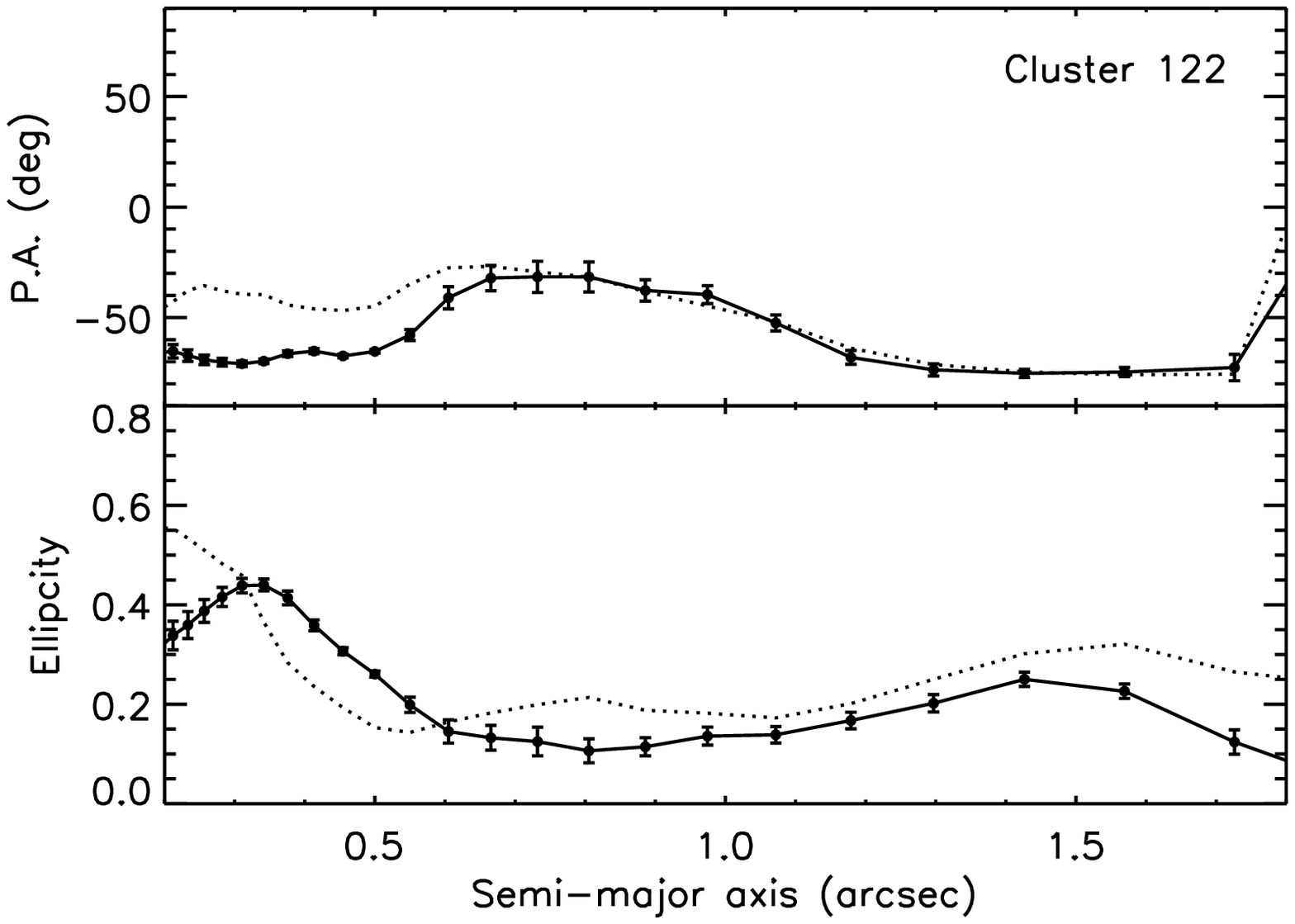} 

\caption{Ellipticities and position angles as a function of semi-major axis for a representative subsample of the total cluster sample. Solid lines correspond to values from F606W images while dotted lines correspond to F814W images. Position angles are counted north through east. The identification name of each cluster correspond with the notation in Paper I.}
\label{ellip_sample}
\end{center}
\end{figure*}

A second pass of ELLIPSE was run to determine ellipticities and position angles (PA), this time fixing each cluster center to the previously calculated values. The ellipticity is a complex quantity that can vary at different radii from the cluster center. To analyze any potential variation, elliptical isophotes were fitted between 0''.1 and the largest measurable semi-major axis distance in both filters. Figure \ref{ellip_sample} shows the ellipticity and PA profiles as a function of semi-major axis for a representative sample of clusters. The arbitrary fluctuations in the ellipticities and PAs at small radii ($<$ 0''.5) are likely to be produced by internal errors in the ELLIPSE algorithm. In the far outer parts of the clusters, the ellipticity and PA are poorly constrained due to the low signal-to-noise ratio. Given these facts, the final ellipticity and PA for each cluster was calculated as the average of the values between 0''.5 and 1''.5 where the quantities are more stable. In most of the cases, the ellipticity is well traced in the different filters. In 12 cases, the faintness of the objects prohibited us from an accurate determination of the ellipticity in both the F606W and F814W filters. For small ellipticities, the PA is not well determined and varies significantly between filters. We chose not to use PA values which had errors exceeding 20$\%$, and hence we have excluded those PA values from the final catalog. Table \ref{Table1} shows the ellipticities and PAs for the star cluster sample identified in Paper I in the F606W and F814W filters. Errors correspond with the standard deviation of the mean.

\begin{figure*}
\begin{center}
\includegraphics[width=0.9\textwidth]{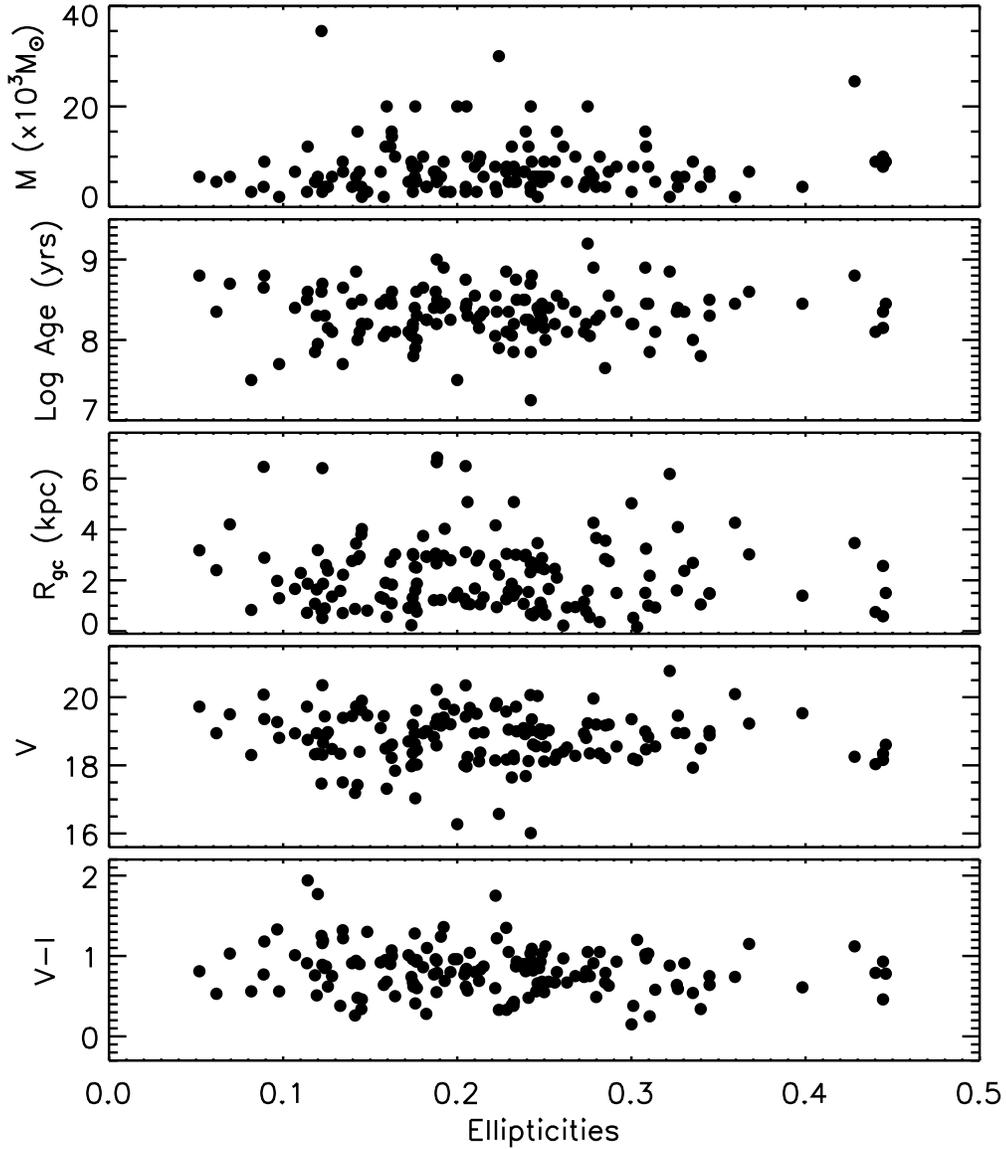}
\caption{Cluster ellipticities versus mass (M), age, galactocentric distance ($R_{gc}$), V magnitude and V-I color. The cluster properties were adopted from Paper I.}
\label{ellip_properties}
\end{center}
\end{figure*}

Some studies have suggested that cluster flattening is generally attributable to cluster rotation effects rather than galactic tides \citep{White1987,Davoust1990}. During the dynamical evolution of these objects, they lose mass and angular momentum. As a consequence, clusters rotate slower and become rounder as they evolve. To investigate the elongation origin of M33 SCs, Figure \ref{ellip_properties}  shows different cluster properties as a function of ellipticity for F606W. The cluster properties were determined and analyzed in Paper I. Ellipticities in M33 clusters do not seem to be driven by any of the cluster properties plotted in Figure \ref{ellip_properties}. 

The distributions of cluster flattening in nearby galaxies are known to be very different \citep[e.g.][]{vandenBergh2008}. Ellipticities in Galactic GCs have been found to correlate inversely with age, suggesting structural changes over time or changes in the initial conditions \citep{Frenk1982}. The correlation of the ellipticity with the age and mass in LMC and SMC clusters are found to be weak \citep{Goodwin1997,Hill2006}. Our M33 sample does not show any clear correlation with age or mass. A luminosity -- ellipticity relation has also been suggested in some nearby galaxies \citep{vandenBergh2008}. While the faintest clusters associated with massive galaxies (MW, M31 and NGC 5128) are flatter than the most luminous star clusters, no evidence for a luminosity -- ellipticity correlation exists in the LMC or SMC. No correlation between magnitude and ellipticity is found in our sample either. \citet{Elson1991} suggest that the high ellipticity observed in many of the young LMC clusters could be due to the presence of subclumps. However, such substructures will be erased as clusters evolve. 

\begin{figure*}
\begin{center}
\includegraphics[width=0.9\textwidth]{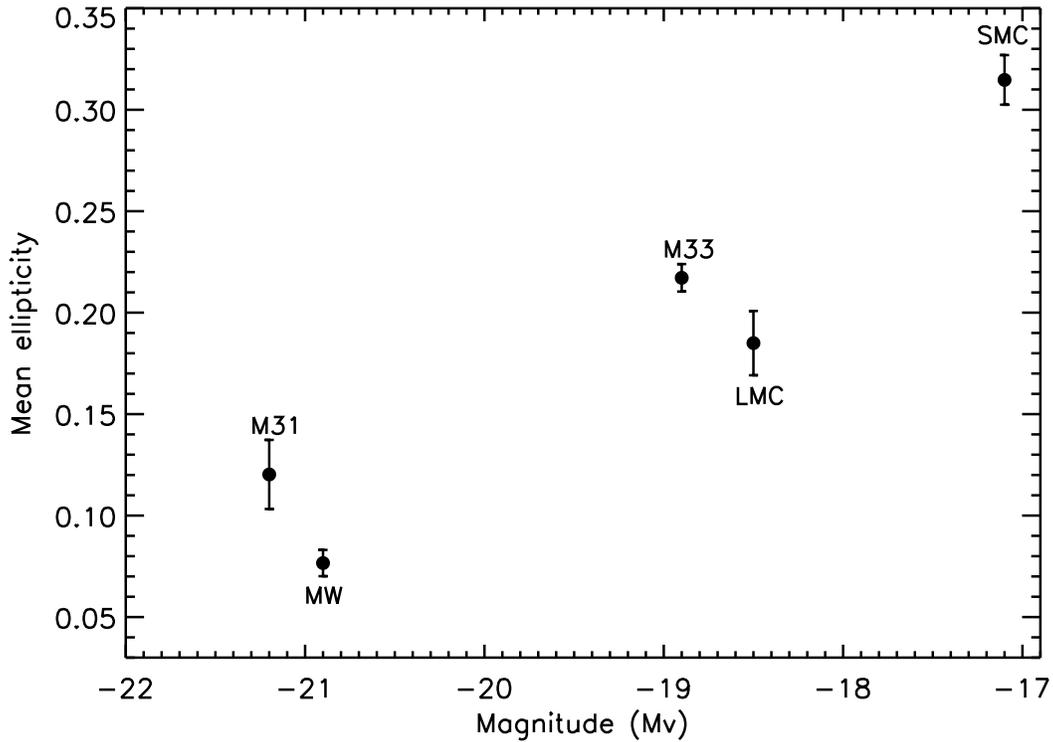}
\caption{Mean ellipticity of star clusters as a function of their host galaxy luminosity.}
\label{mean_ellip1}
\end{center}
\end{figure*}

The mean ellipticity of star clusters in the MW/M31 is smaller than that of the LMC, which is in turn smaller than the mean ellipticity of star clusters in the SMC.  It has been suggested that this is correlated with the masses of these galaxies \citep{Geisler1980,Staneva1996}. \citet{Goodwin1997} suggests the  tidal field strength of the parent galaxy as the dominant factor driving the differences between the LMC and the MW star clusters. Tidal forces would destroy the initial anisotropic velocity dispersions of a cluster, modifying its initial shape. This effect would be much less significant in the less massive galaxies so the tidal field in the LMC would not be able to modify significantly the shapes of its clusters maintaining their initial triaxiality. This scenario would also explain the even higher ellipticities observed in the SMC.  Figure \ref{mean_ellip1} shows the mean ellipticity of different star cluster systems as a function of the luminosity of their host galaxy. The mean ellipticities were obtained from \citet{Harris1996} for the MW, \citet{Barmby2002,Barmby2007} for M31, \citet{vandenBergh2008} for the LMC and \citet{Hill2006} for the SMC. The mean ellipticity for clusters in M33 corresponds to the mean value from Table 1 from F606W band. The error of the data corresponds with the standard deviation. The luminosities for each galaxy are adopted from \citet{vandenBergh1999}. With a mean ellipticity of $e$=0.2, M33 clusters are, on average, more flattened than those of the MW or M31 and more similar to clusters in the MC in agreement with this scenario. 

\begin{figure*}
\begin{center}
\includegraphics[width=0.9\textwidth]{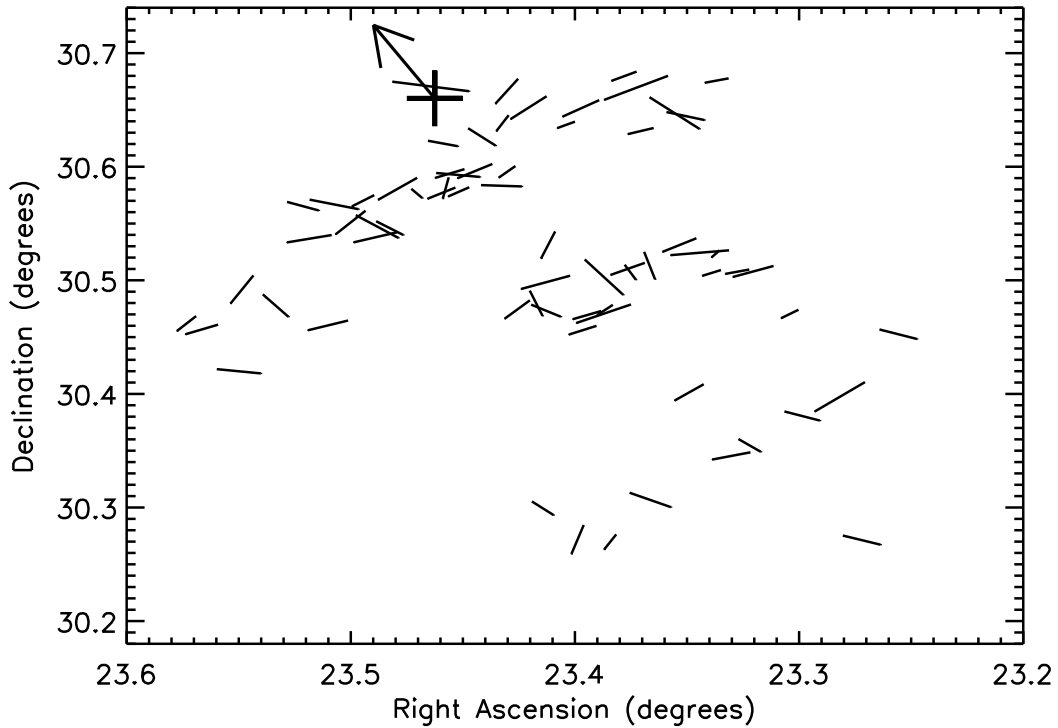}
\caption{Cluster elongations and orientations represented in the RA-Dec coordinate plane. North is up and east is to the left. The vector sizes are correlated with the ellipticity obtained in this study. The cross corresponds with the center of M33, with the arrow marking the position angle (23$^{o}$) of the galaxy major axis \citep{Regan1994}.}
\label{pa_orientation}
\end{center}
\end{figure*}

\begin{figure*}
\begin{center}
\includegraphics[width=0.9\textwidth]{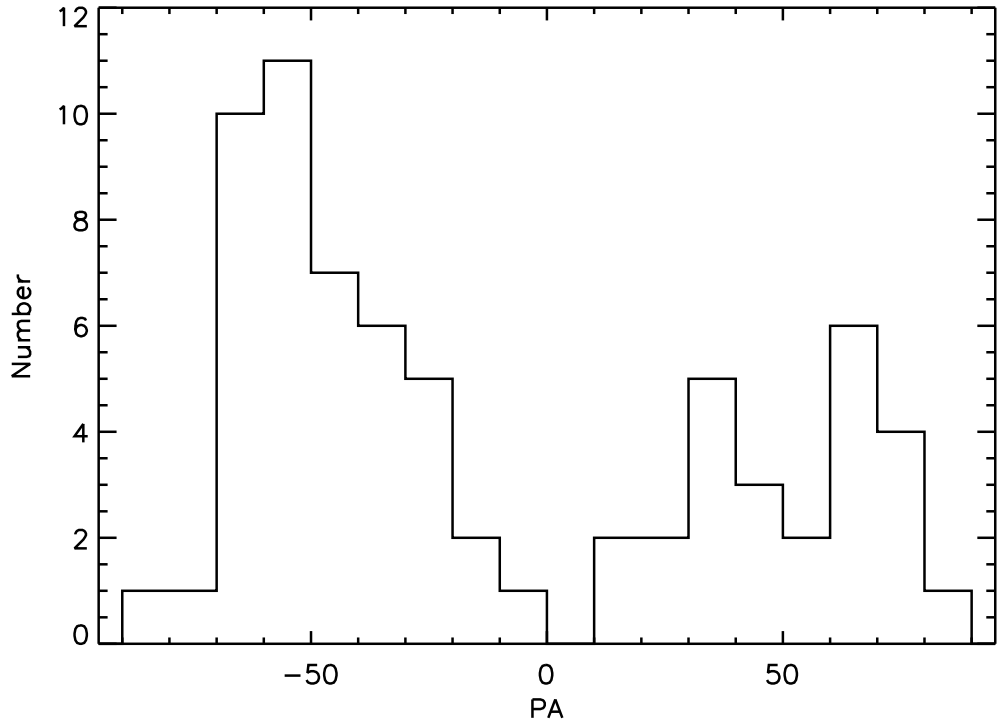}
\caption{Distribution of the position angles (PA) of our sample.}
\label{pa_distribution}
\end{center}
\end{figure*}

On the other hand, tidal fields are also be able to stretch clusters and make them more elongated. If tidal forces have a dominant effect on the elongation of M33 clusters, one would expect a preferred orientation of the SC position angles towards the location of the force. Figure \ref{pa_orientation} summarizes the elongation and orientation of the clusters with respect to the galaxy. There is no clear trend in the orientation vectors toward the galactic center, however the orientation of the clusters tend to have a preferred direction. In particular, the position angle distribution in Figure \ref{pa_distribution} shows a bimodality with a strong peak at -- 55$^{o}$. This northwest direction points toward M31, which is located at a position angle of about -- 40$^{o}$ with respect to the center of M33. Recent evidence suggests that a close encounter between M33 and M31 could have affected the properties of the M33 disk \citep{Putman2009,McConnachie2009}. Due to the small galactocentric distance of these clusters ($R_{gc}$ $<$ 7 Kpc), the idea of M31 as the source of the elongation seems unlikely. Analysis of the PAs show no correlation with age, mass, distance or magnitude of the clusters. No correlation has been found with the spiral arm pattern or with circular orbits around the galaxy. The elongation of the clusters are not aligned with the x-y pixel coordinate system of each set of images, ruling out a possible artifact related to cluster orientation. The significance of the preferred orientation of the M33 clusters remains open and further investigation of this phenomenon will require precise PAs of a bigger sample of clusters at different locations in the galaxy.

\section{Surface Brightness Profiles}

 As mentioned before, we have derived surface brightness profiles rather than surface number density profiles. The ELLIPSE task was again run to obtain F606W and F814W surface brightness profiles. This time the fixed, zero-ellipticity isophote mode was required because we chose to fit circular models.  We have followed the prescription in \citet{Barmby2007} to transform the raw output values from ELLIPSE to calibrated surface brightnesses on the Vegamag system. We have first converted image counts (counts pixel$^{-1}$) into luminosity density in $L_{\odot}pc^{-2}$. We have worked in terms of linear intensity and finally transform  the results to surface brightness magnitude units ($\mu$/mag arcsec$^{-2}$). The profiles extended out to R=14'' (63pc) in both filters. Figure \ref{Fig5} shows the cluster surface brightness profiles where the error bars are estimations obtained by ELLIPSE from the isophotal intensities. 

\subsection{Irregular Profiles}

Visual inspection of the surface brightness profiles shows that a significant number of  cluster profiles do not decrease smoothly as expected, exhibiting irregularities such as bumps, dips or sharp edges. These irregularities appear in both filters. Figure \ref{Fig5} shows examples of more traditional profiles versus Figure \ref{Fig6} that shows examples of cluster profiles with different irregularities.  For a detailed inspection of the irregualr profiles, we have overplotted 3 concentric apertures, 0.1'', 0.5'' and 1.0'', on the F606W images in Figure \ref{Fig6}. More than half of our sample exhibits some level of anomaly. Some of these features could be the result of statistical fluctuations or due to a few luminous stars. In fact the oversampling of our data could magnify some features in the profiles, conferring an occasional peak due to a very bright star, the degree of a pronounced bump. However,  sub-clumps or a deficit of stars are also prominent in a significant number of clusters (e.g. clusters 12, 24, 71), and appear to reflect real structures within the clusters. 

Similar anomalies have been identified among LMC and SMC star cluster profiles \citep{Mackey2003b,Hill2006, Werchan2011}. They detect, in several surface brightness profiles, systematic deviations from the analytical models that they attribute to a lack of central concentration. They refer to these clusters as ``ring'' clusters and suggest that these anomalies could be the result of dynamical evolution. Some of the clusters could be in the process of dissolution or have had insufficient time to dynamically relax. Variations in the mass to light ratio could also cause departures from the analytical models. \cite{Elson1991} reports bumps, sharp ``shoulders'', and central dips in the profiles of LMC clusters with an age range similar than our sample and interpret these anomalies as consequences of the initial conditions or signatures of merging sub-condensations. As the clusters evolve, the substructure will be erased, but we do not see any correlation between the presence of these anomalies and any other parameters. In particular, objects with irregularities are observed along the entire range of ages with no preference at any particular stage of evolution. Whether any physical significance can be attached to these irregularities is uncertain.

\begin{figure*}
\begin{center}
\includegraphics[width=0.49\textwidth]{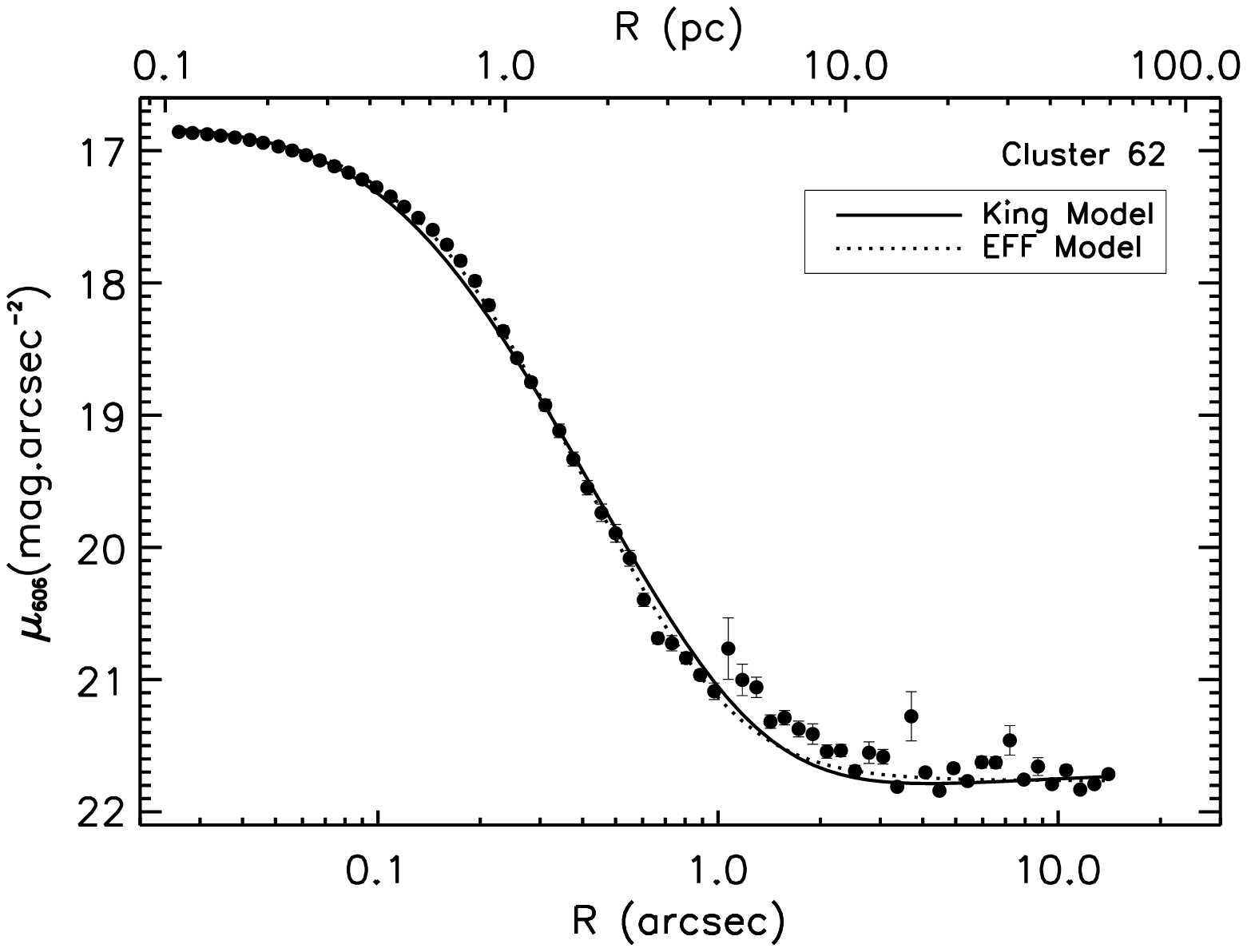} 
\includegraphics[width=0.49\textwidth]{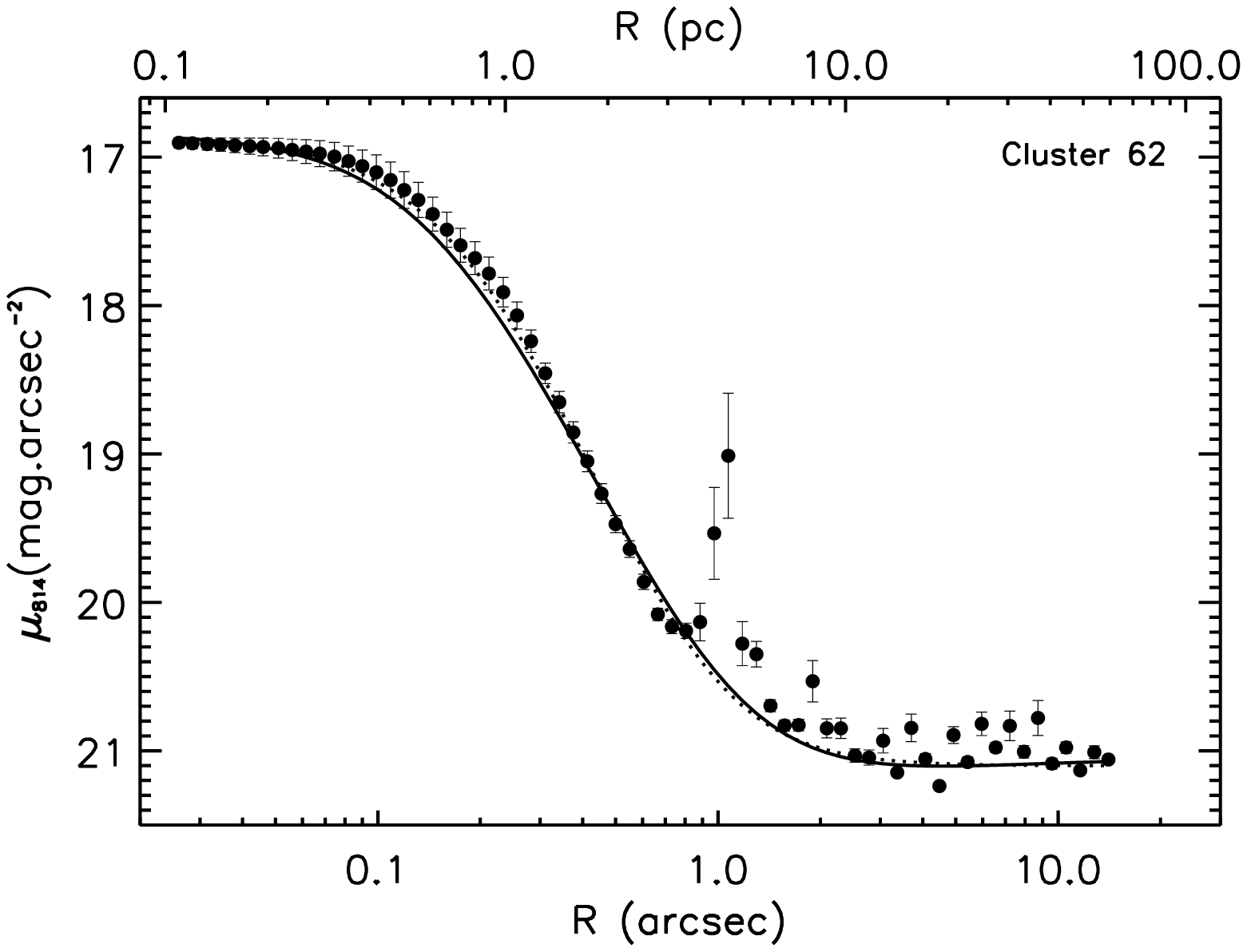} 
\includegraphics[width=0.49\textwidth]{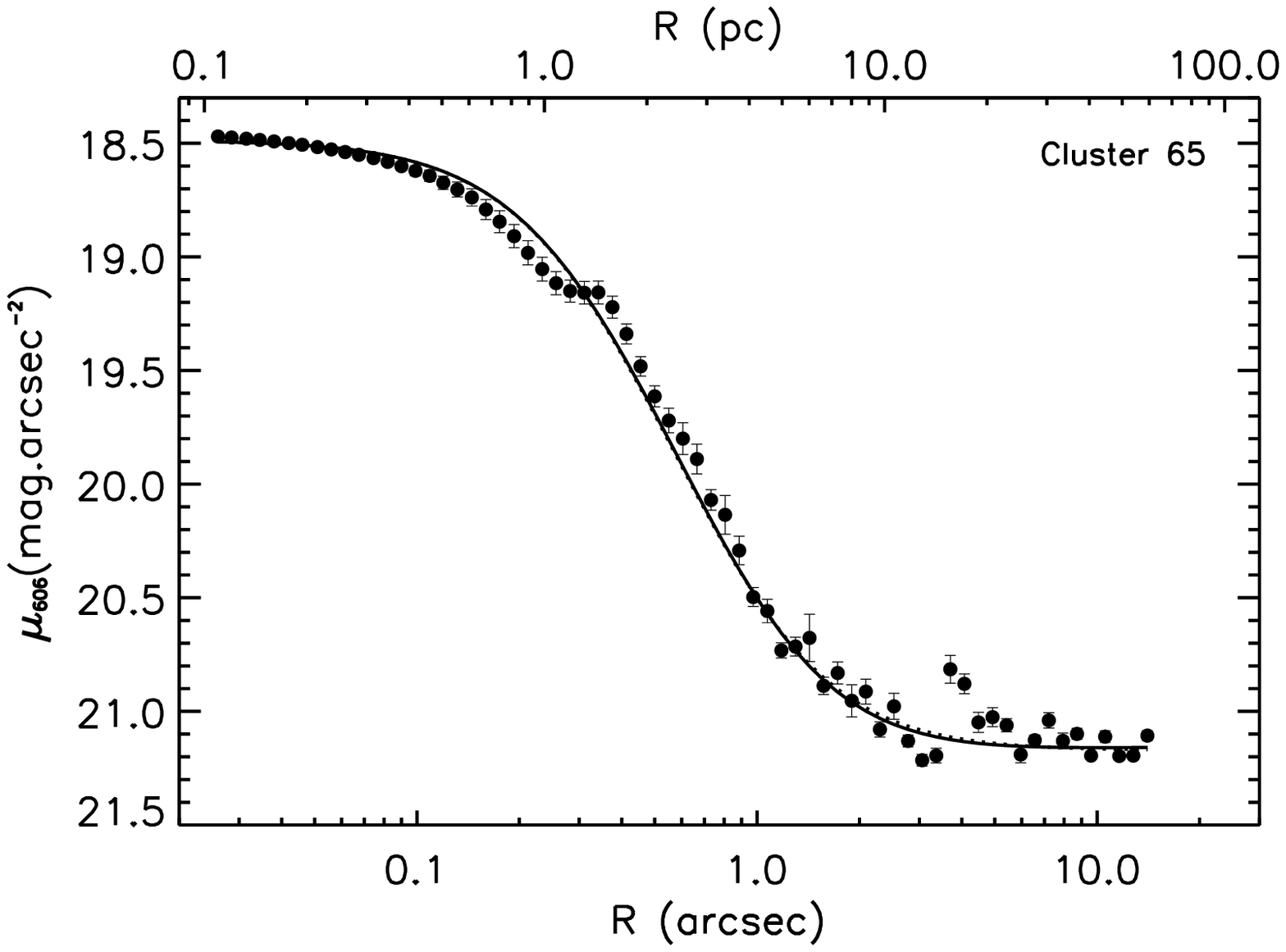} 
\includegraphics[width=0.49\textwidth]{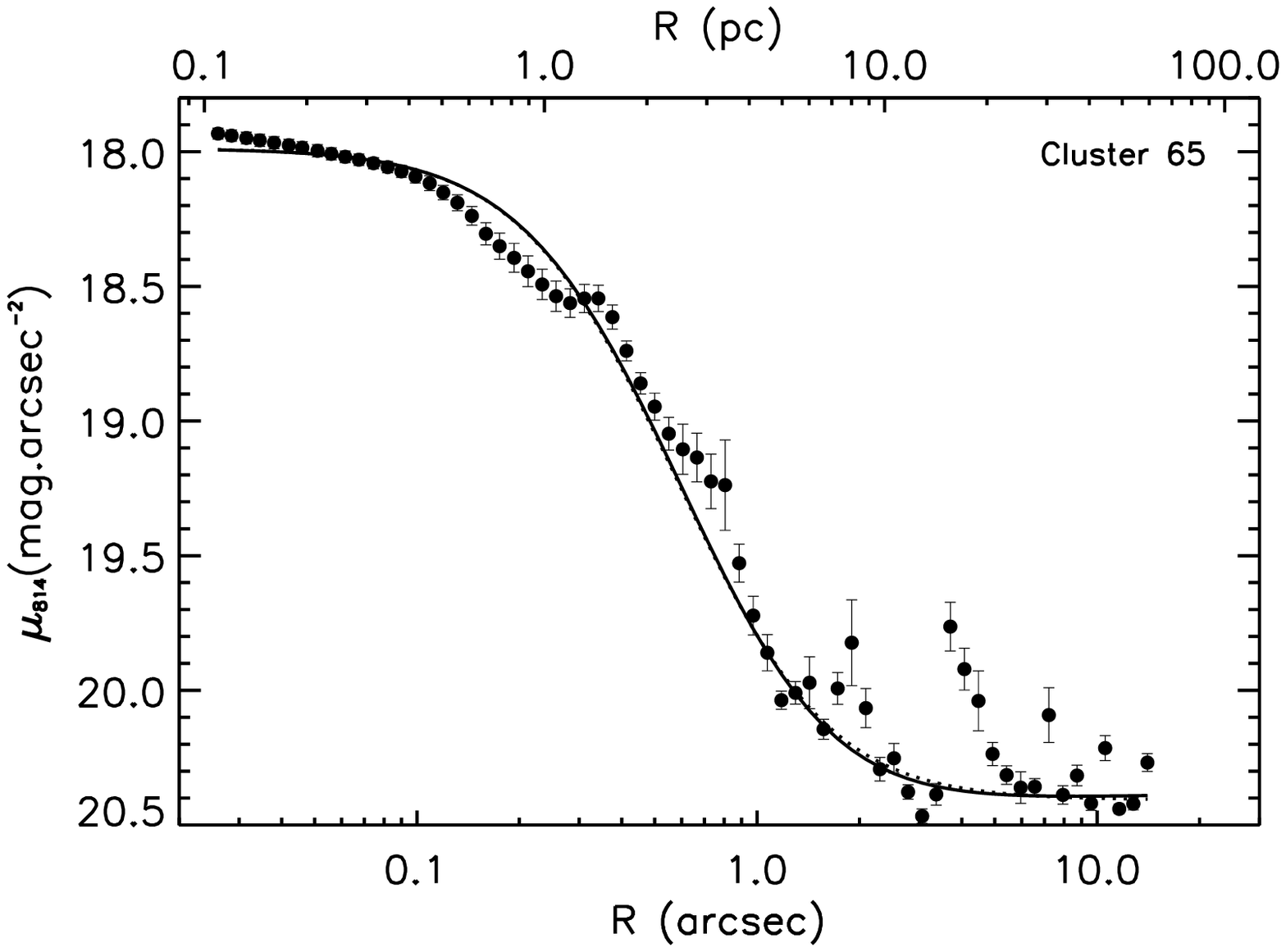}
\includegraphics[width=0.49\textwidth]{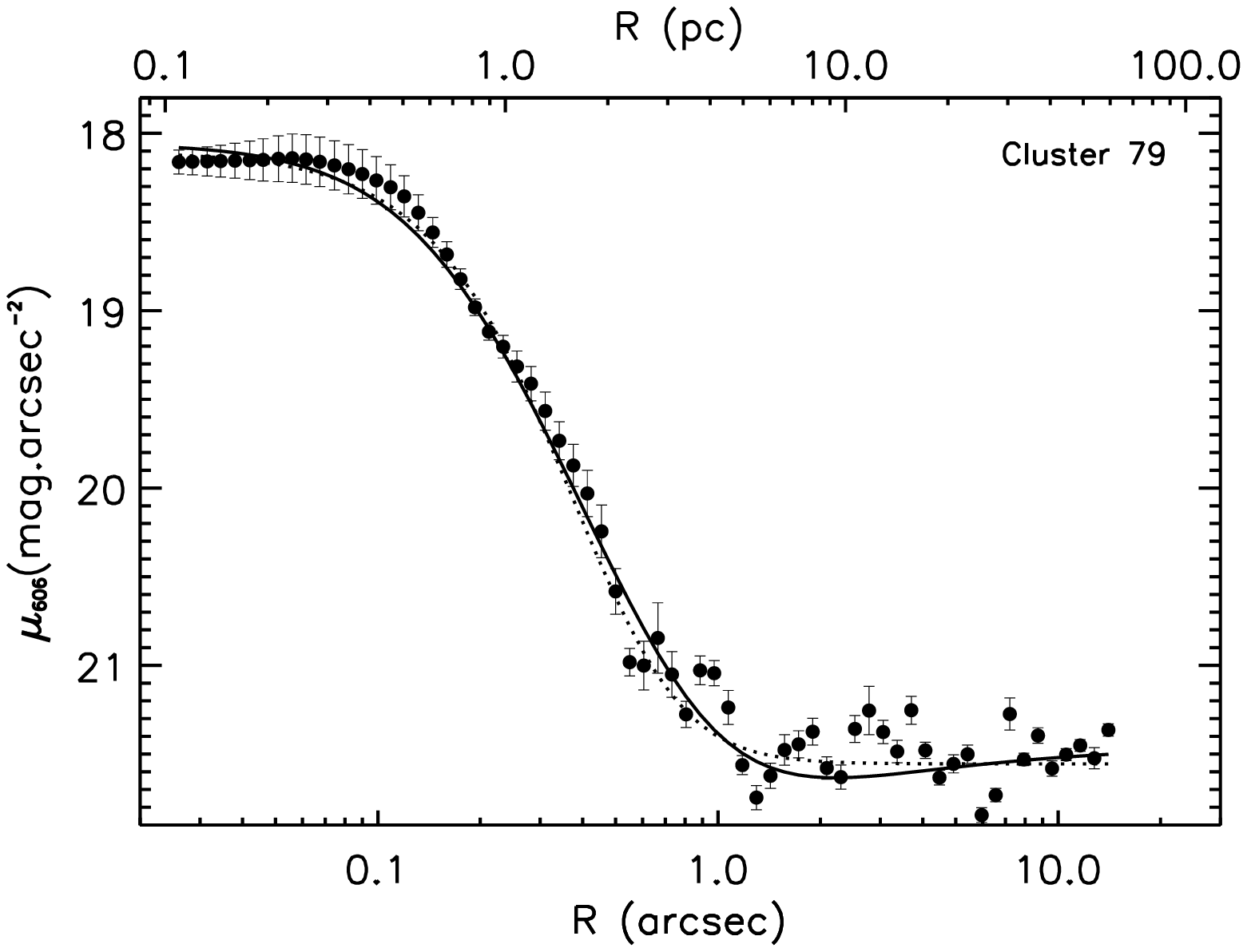} 
\includegraphics[width=0.49\textwidth]{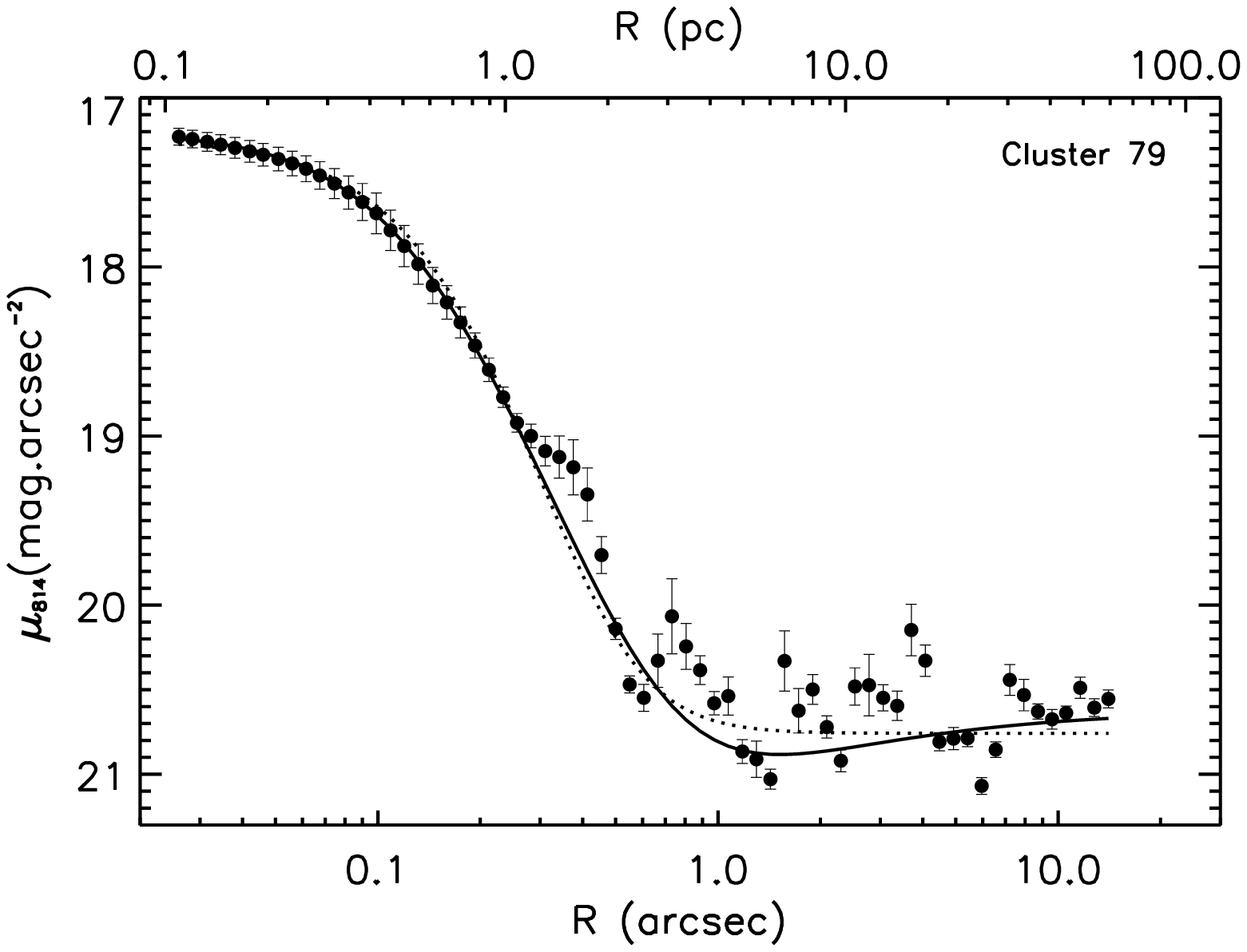}
\caption{\label{Fig5}
Surface brightness profiles and model fits to a representative sample of clusters. Left panels correspond with F606W profiles and right panels with F814W. The solid line in each panel represents the best-fit King model while the dotted line represents the best-fit EFF model.
}
\end{center}
\end{figure*}
\addtocounter{figure}{-1}
\begin{figure*}
\begin{center}
\includegraphics[width=0.49\textwidth]{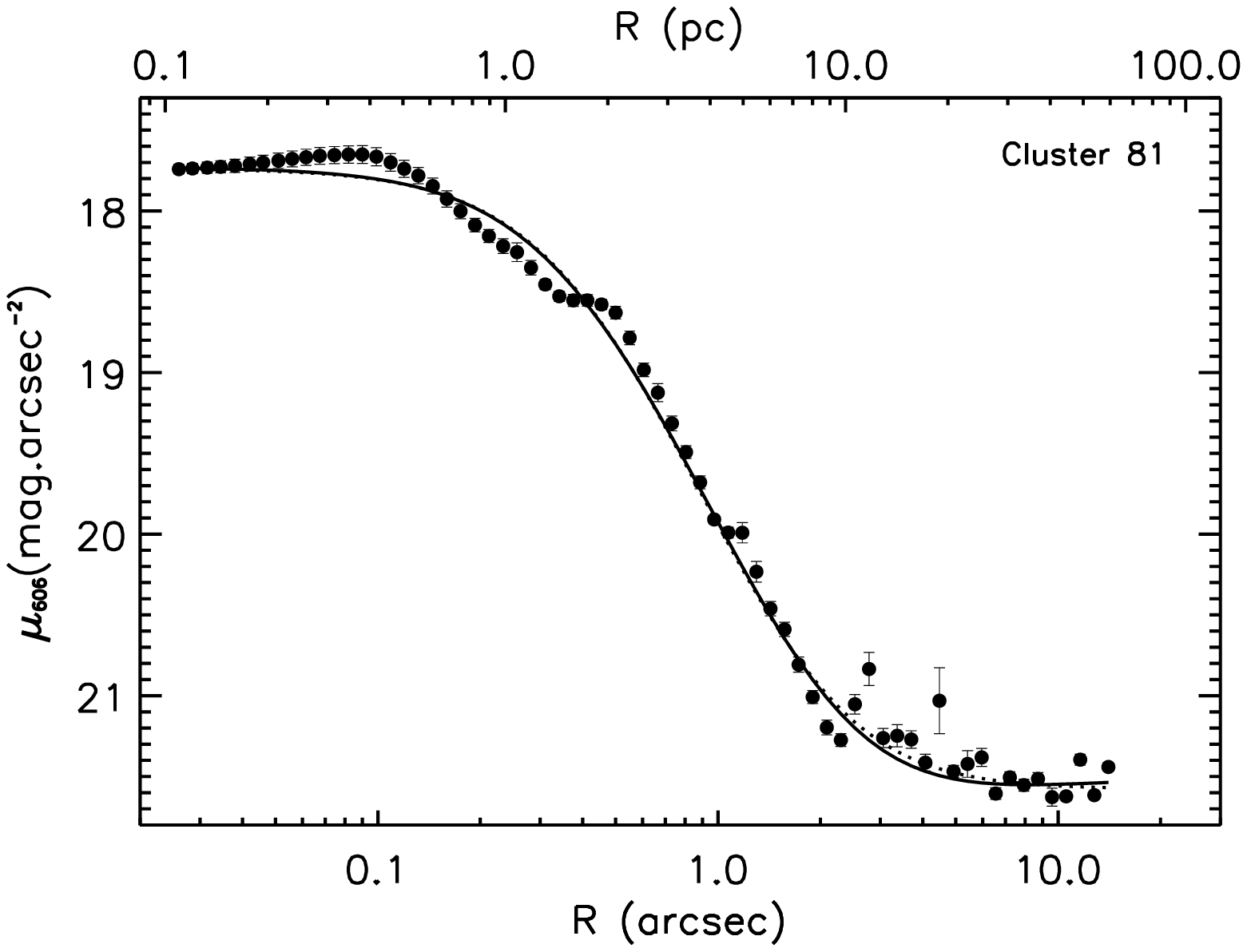} 
\includegraphics[width=0.49\textwidth]{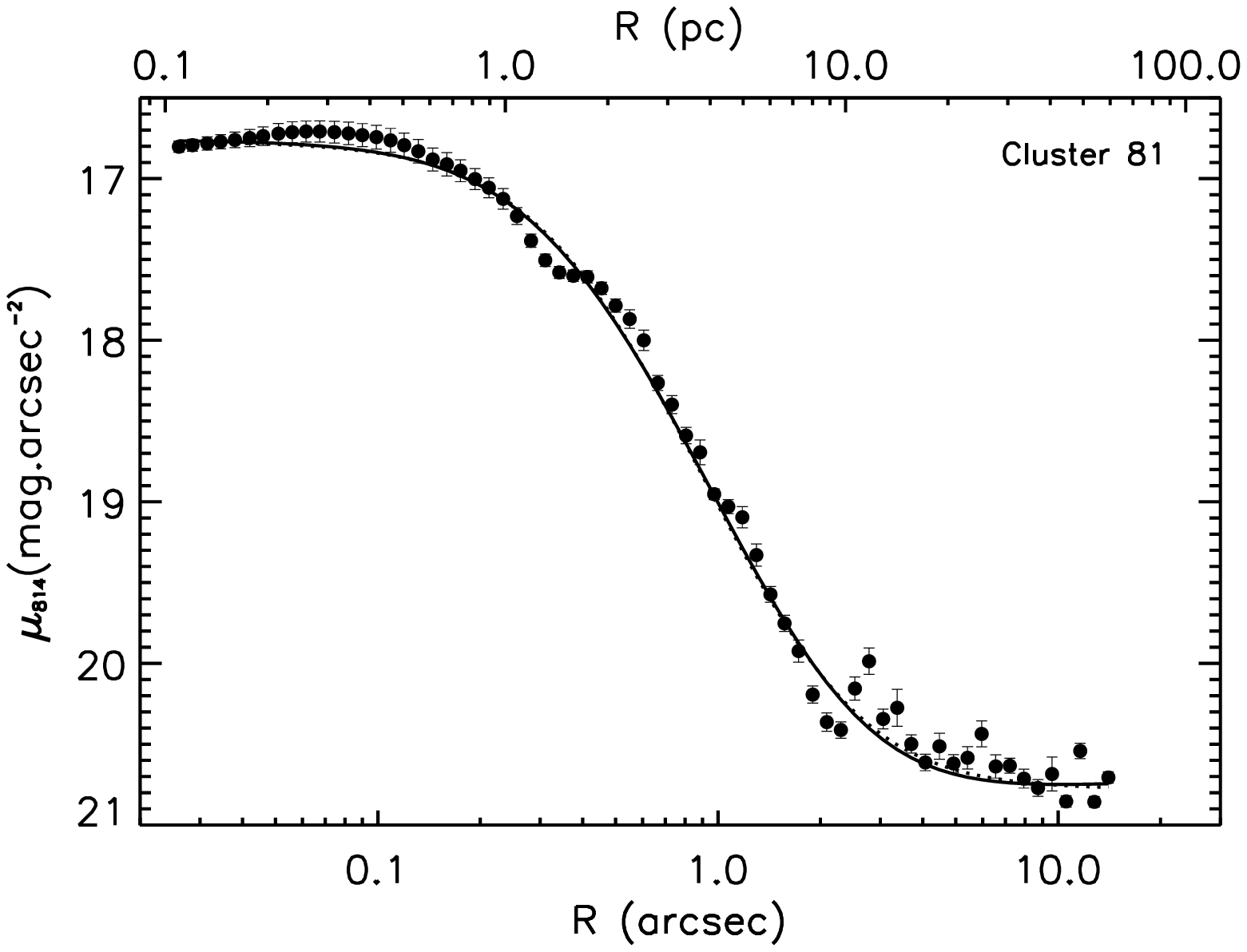} 
\includegraphics[width=0.49\textwidth]{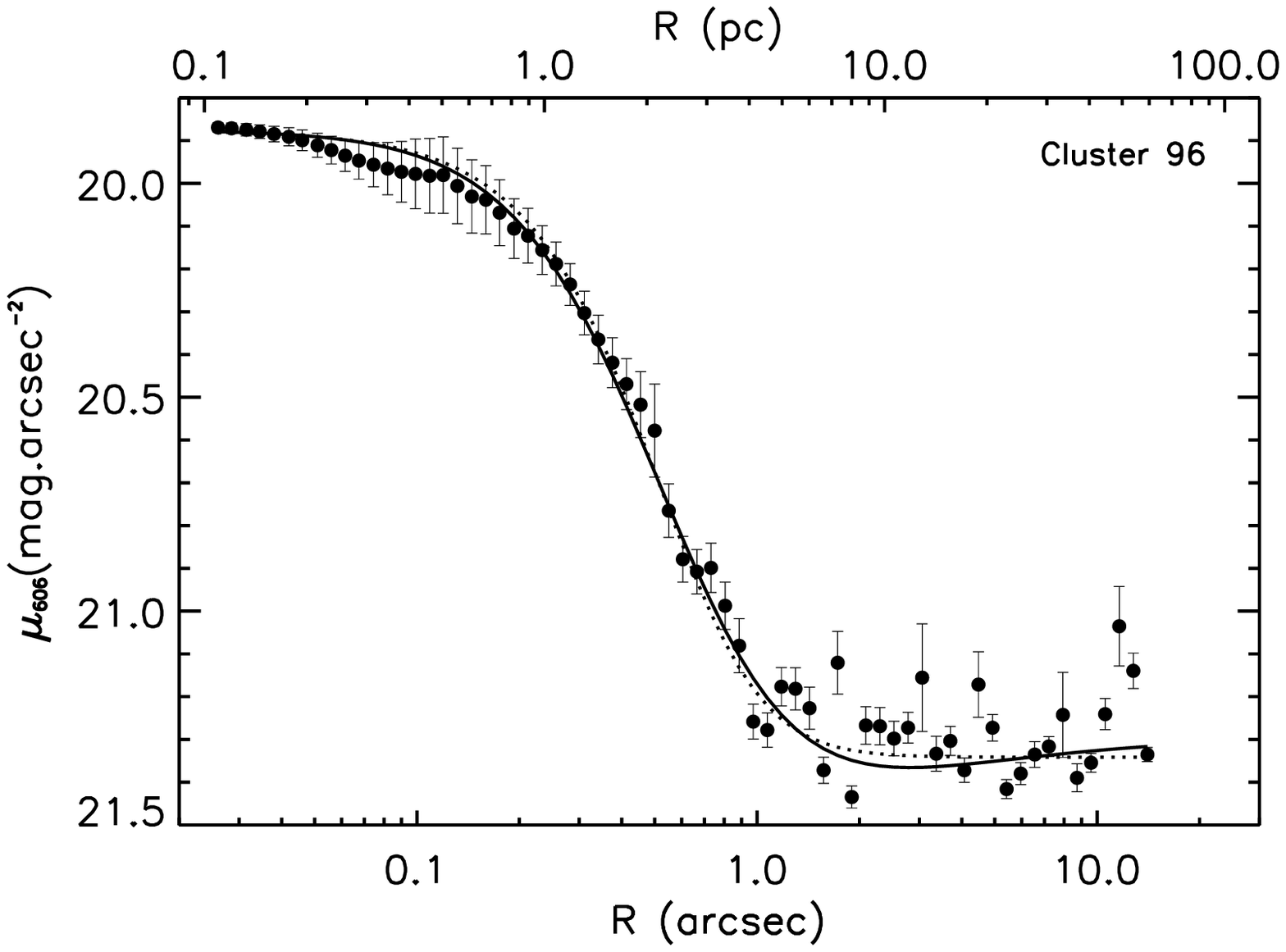} 
\includegraphics[width=0.49\textwidth]{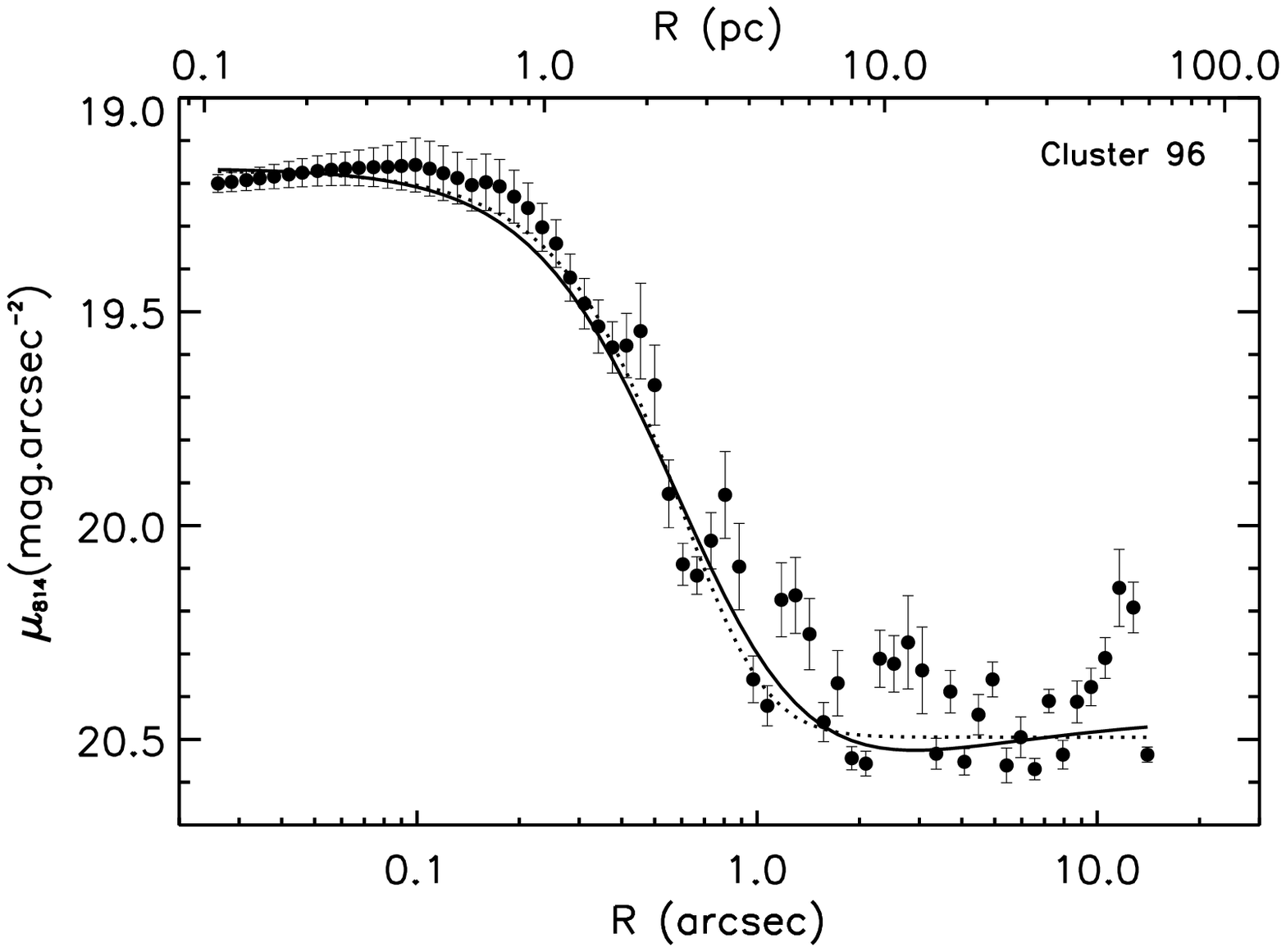} 
\includegraphics[width=0.49\textwidth]{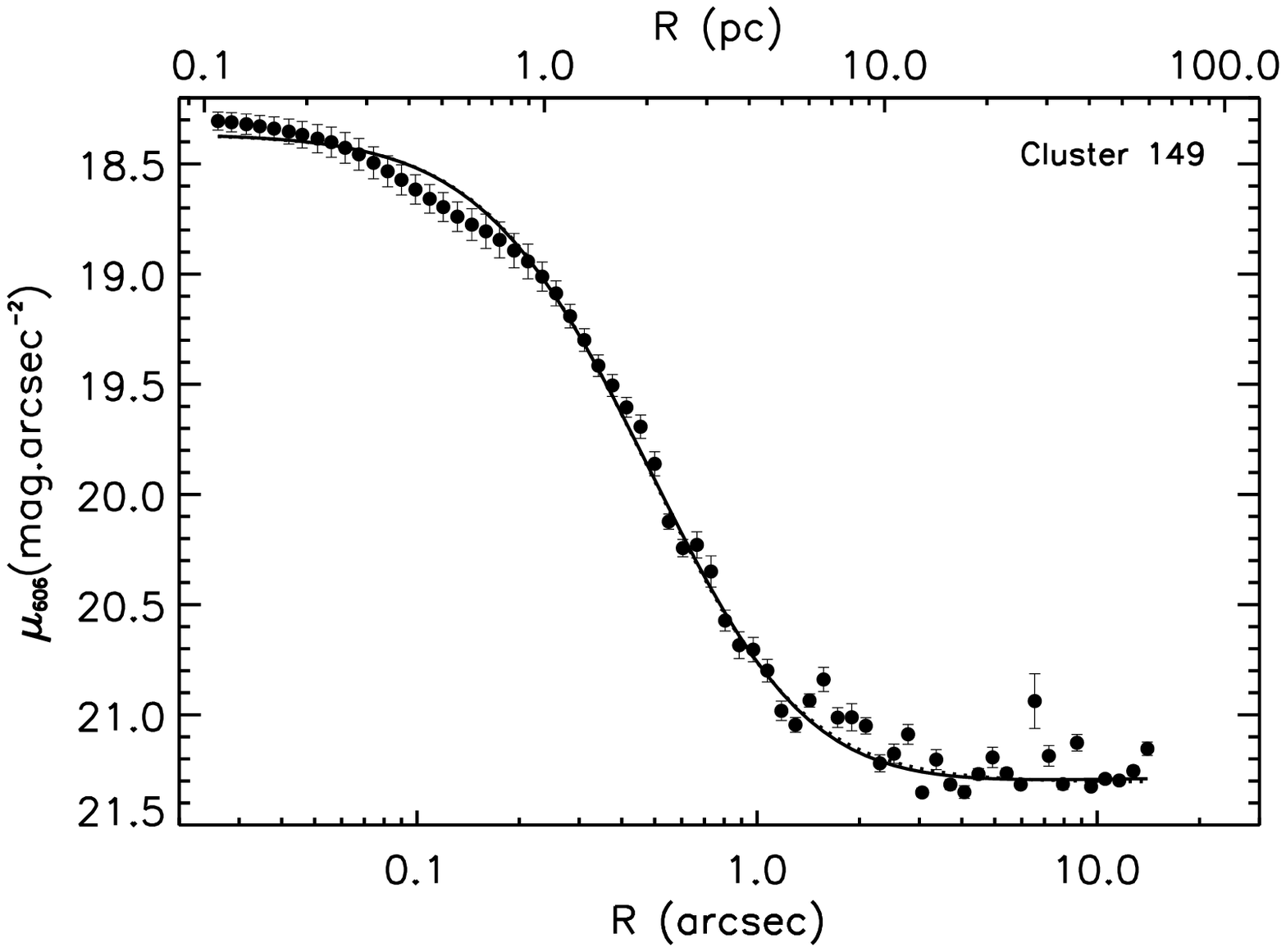} 
\includegraphics[width=0.49\textwidth]{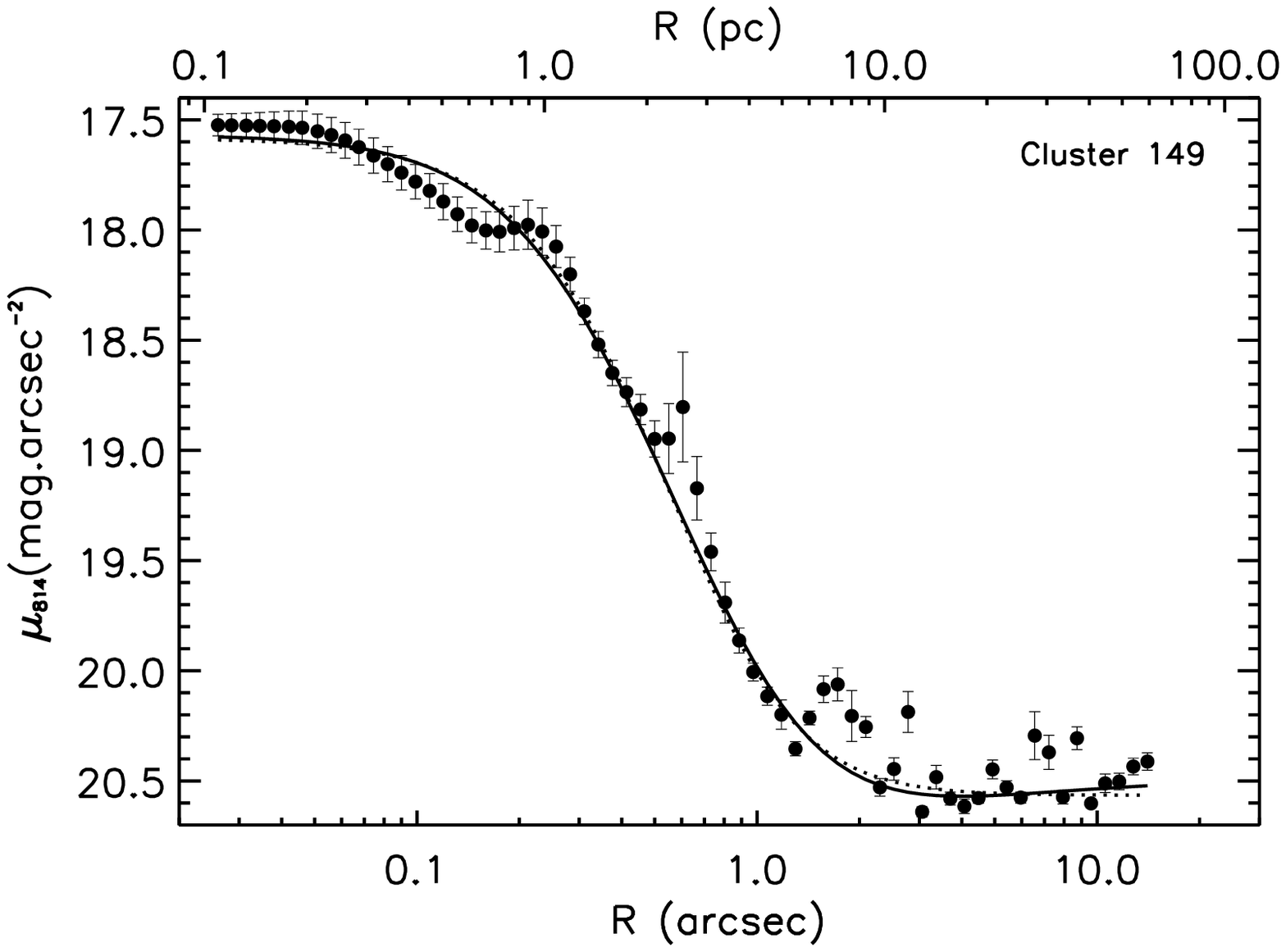} 
\caption{Continued}
\label{before}
\end{center}
\end{figure*}

\begin{figure*}
\begin{center}
\includegraphics[width=0.49\textwidth]{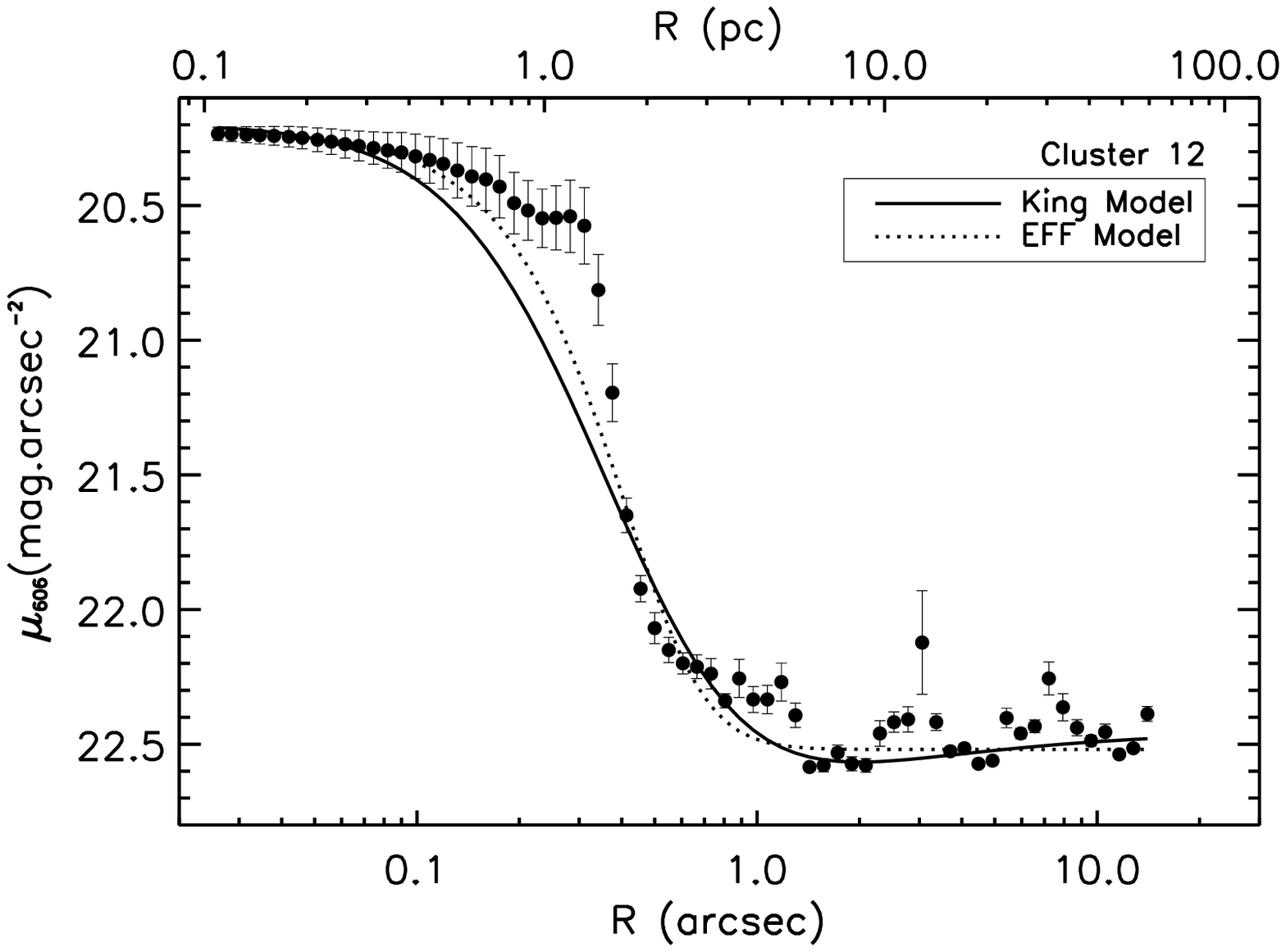} 
\includegraphics[width=0.49\textwidth]{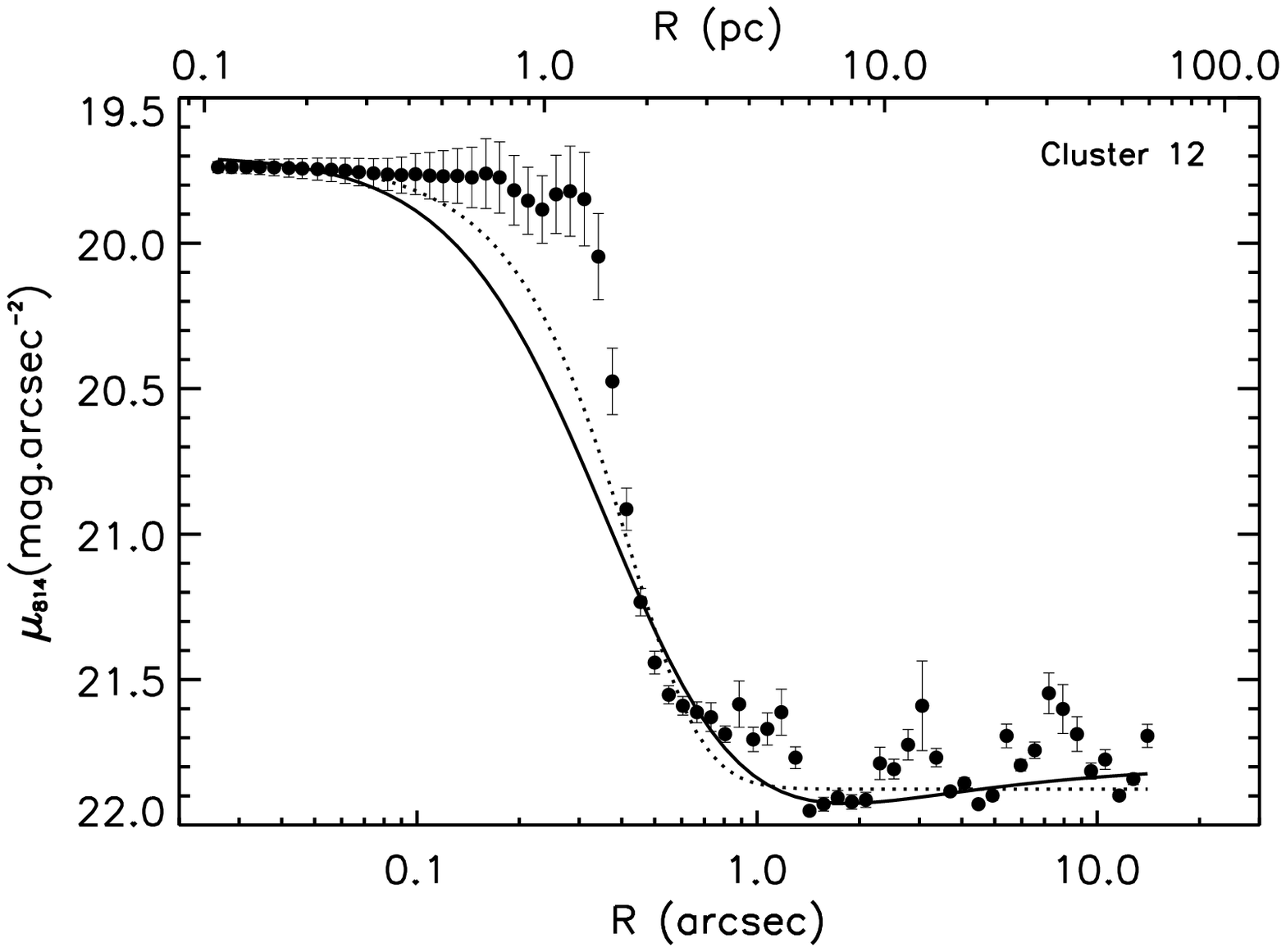} 
\includegraphics[width=0.49\textwidth]{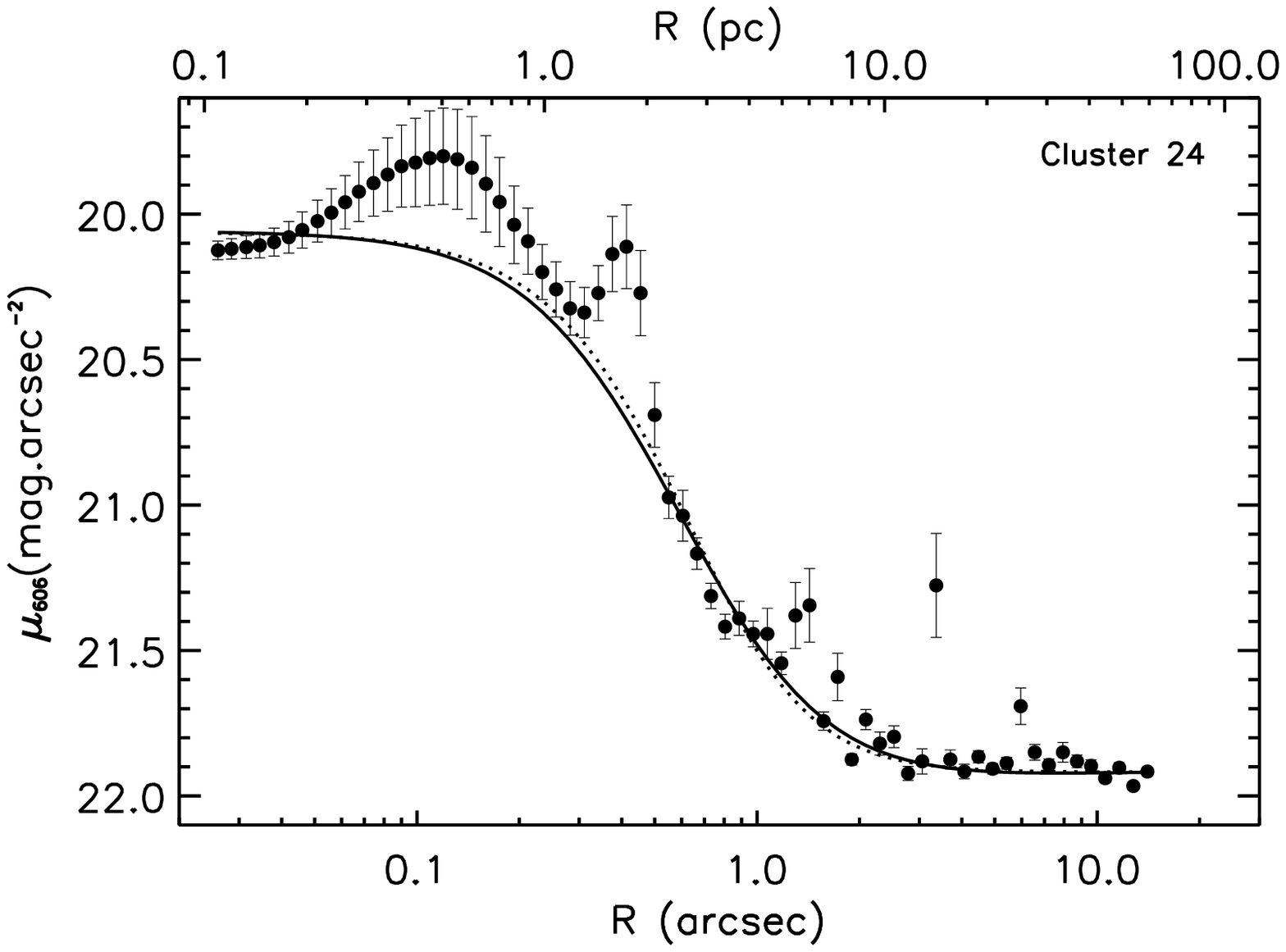} 
\includegraphics[width=0.49\textwidth]{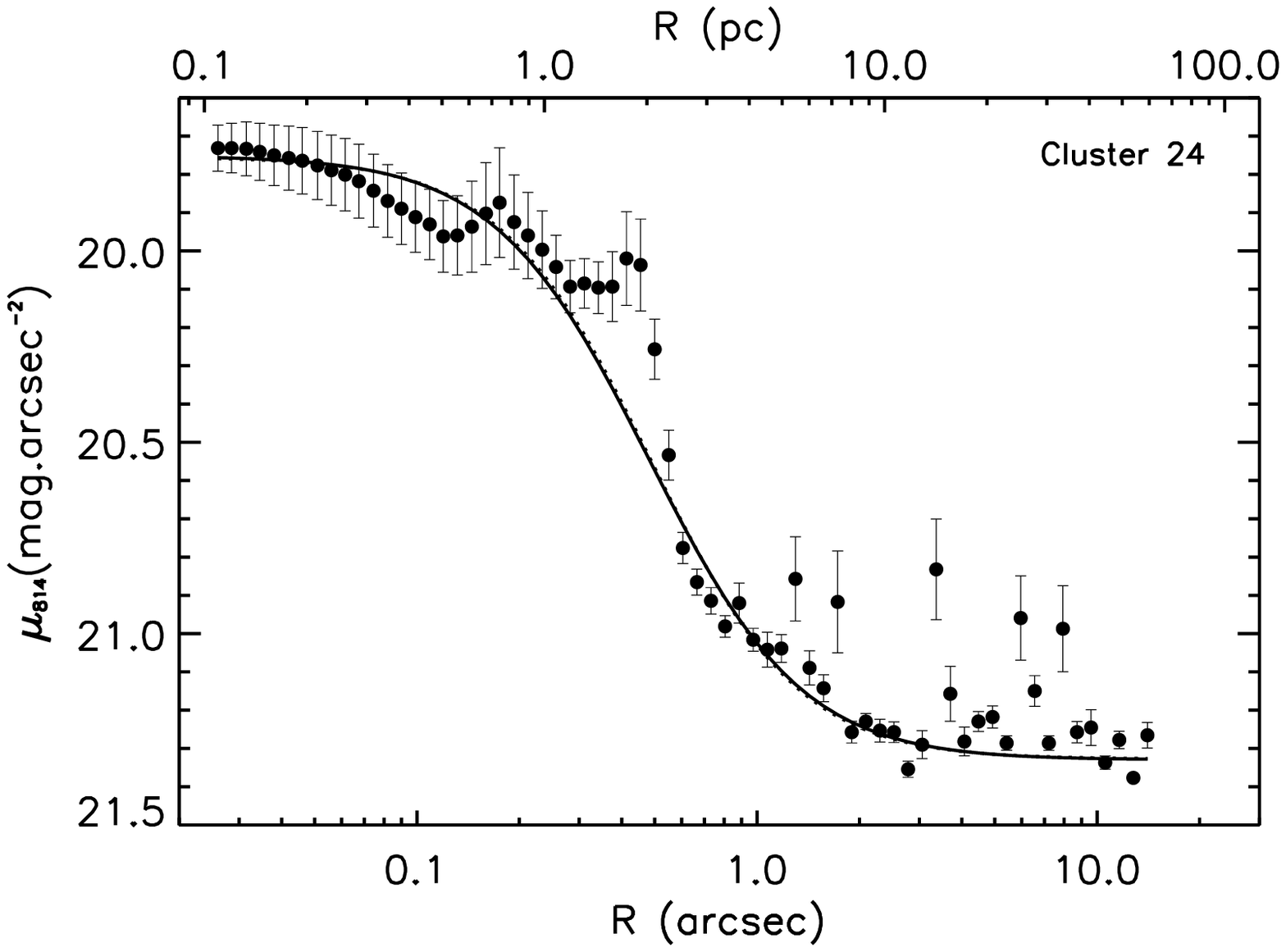}
\includegraphics[width=0.49\textwidth]{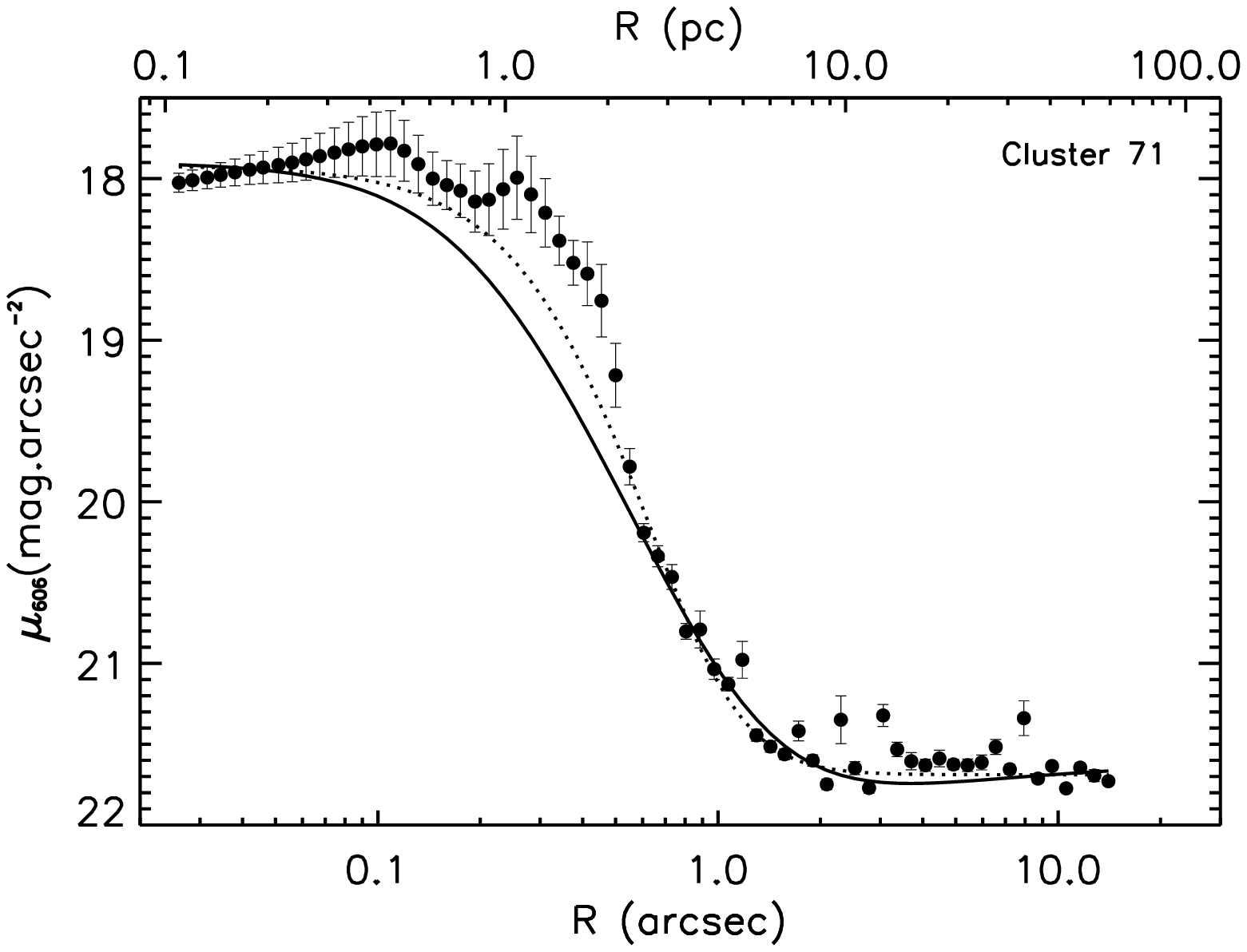} 
\includegraphics[width=0.49\textwidth]{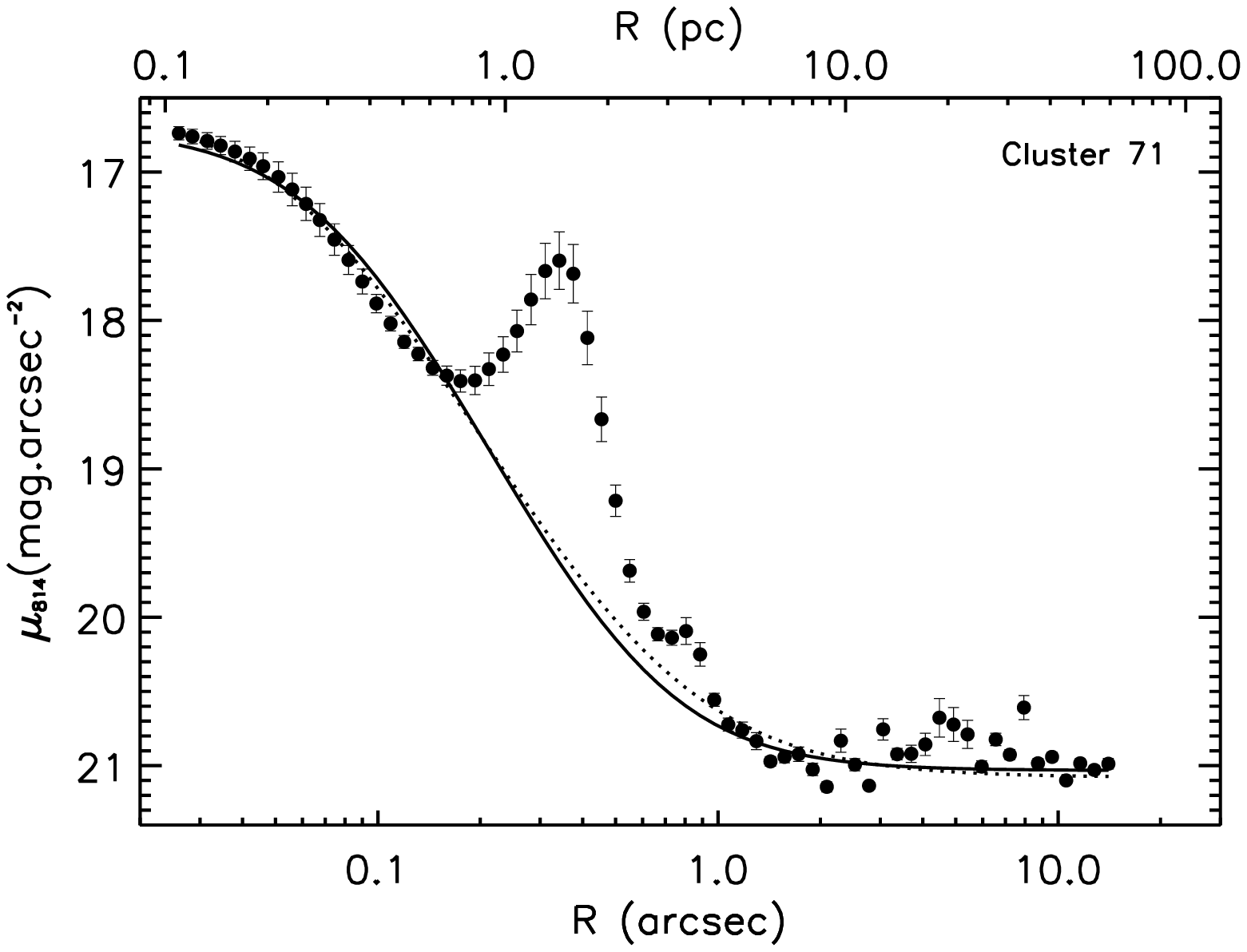} 

\includegraphics[width=0.2\textwidth]{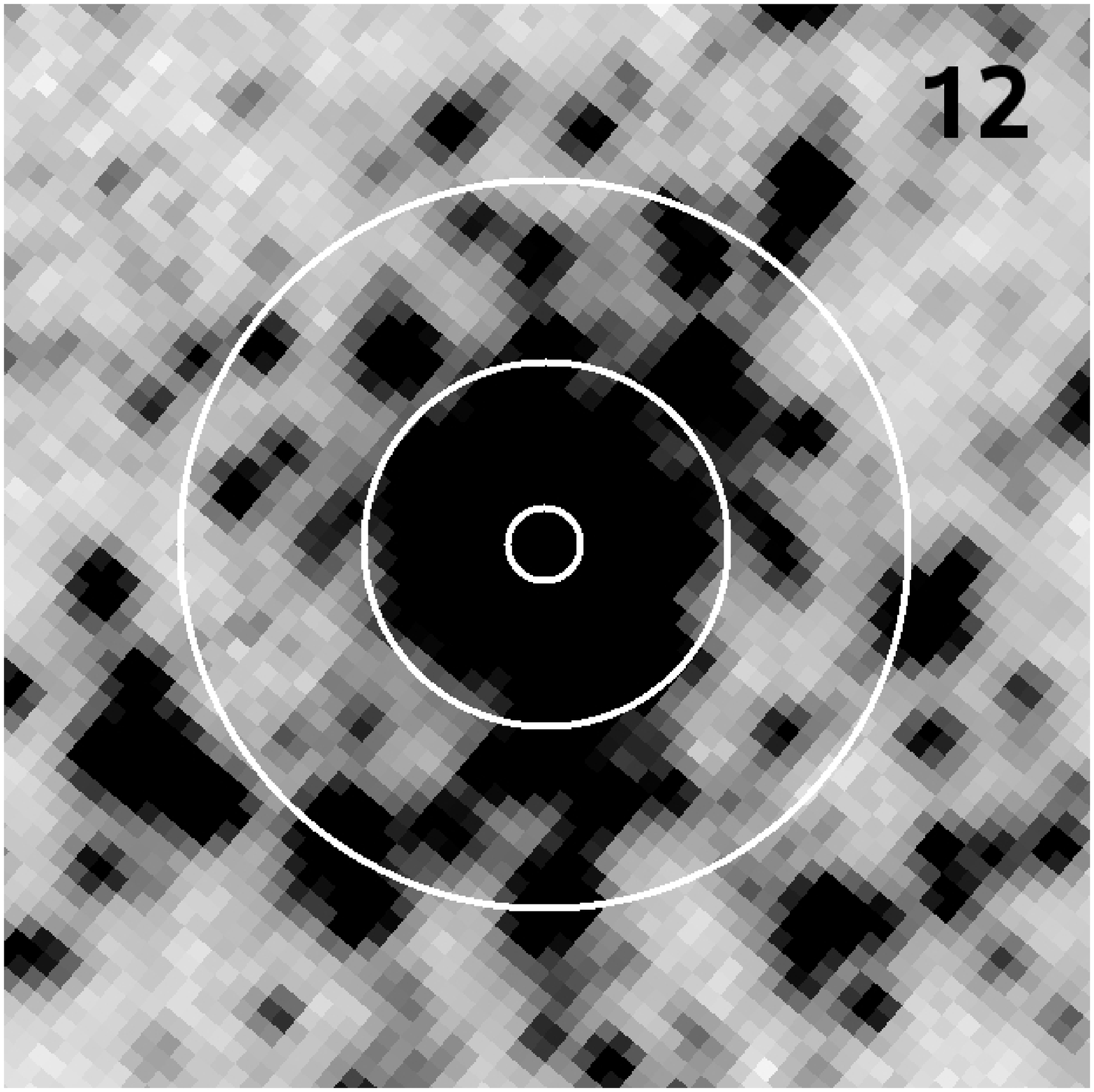}
\hspace{9pt}
\includegraphics[width=0.2\textwidth]{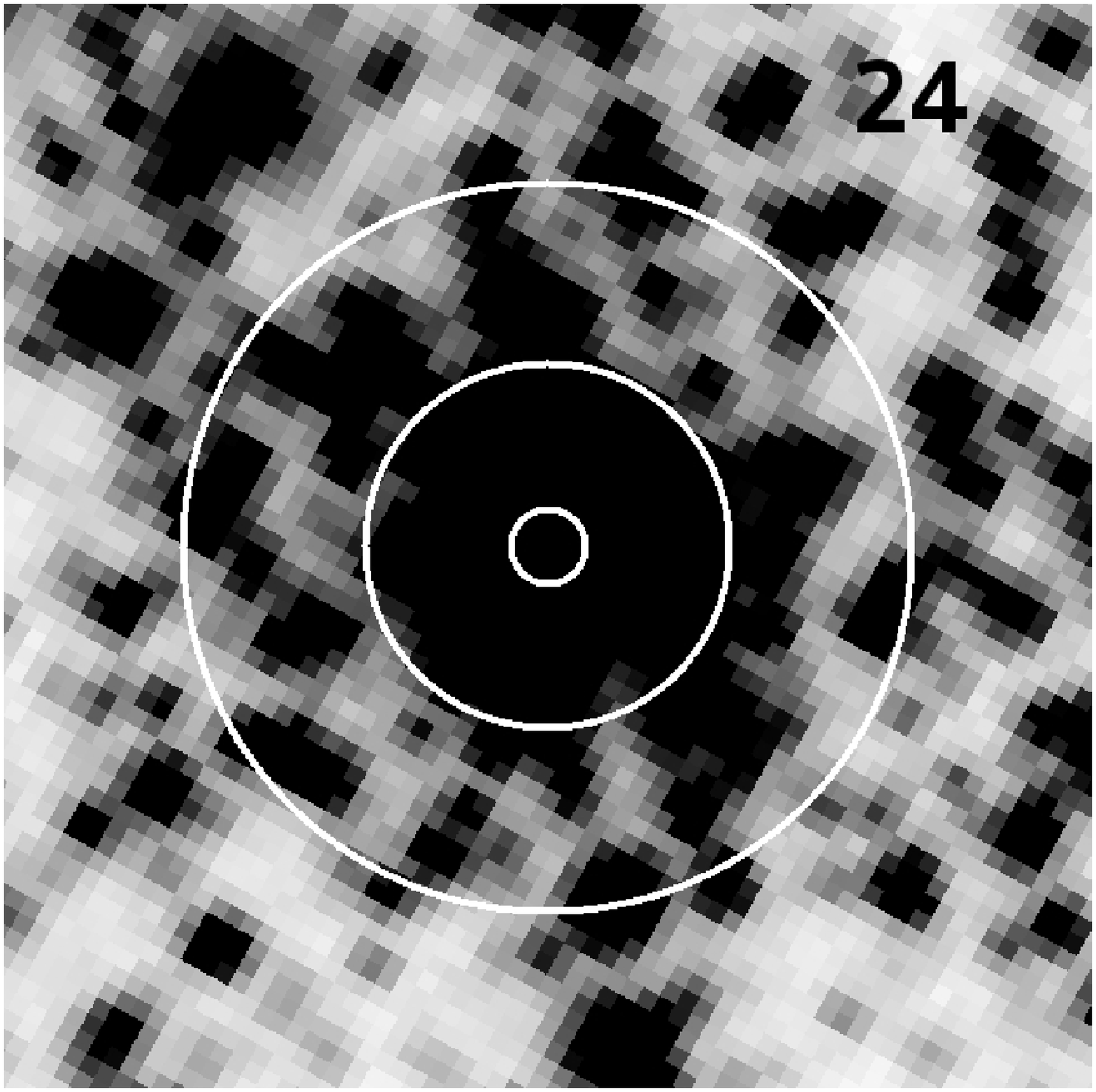}
\hspace{9pt}
\includegraphics[width=0.2\textwidth]{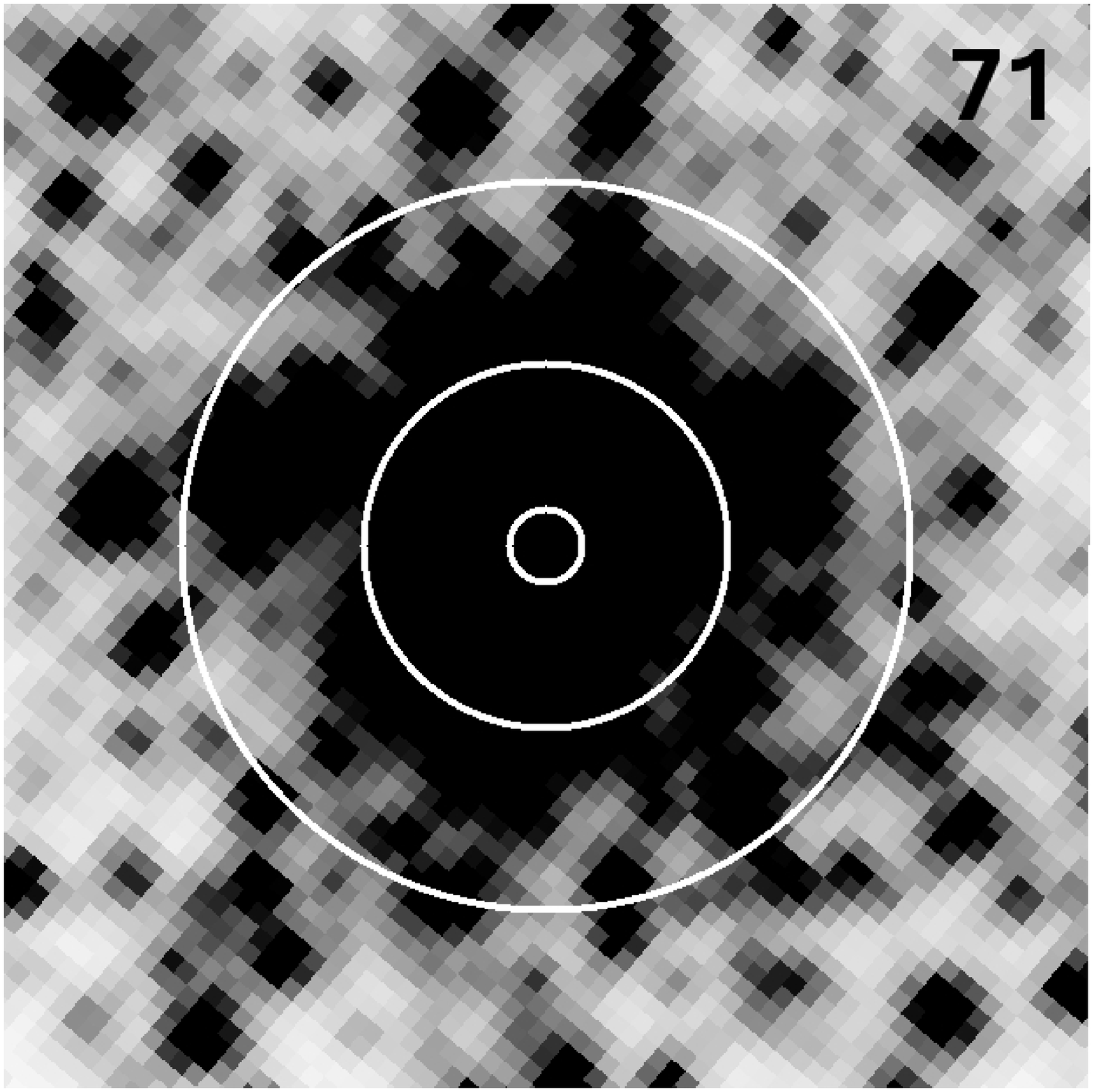}
\caption{Same as Figure \ref{before} but for a group of clusters showing irregularities in their profiles. Each row corresponds with a different cluster with left panel showing the F606W profile and right panel the F814W profile. The bottom row shows a 3''x 3'' F606W image of the 3 clusters. Each image is shown with the same gray-scale with north up and east to the left. For visual guiding, the 3 concentric apertures correspond with radii of 0.1'', 0.5'' and 1''.0.}
\label{Fig6}
\end{center}
\end{figure*}

\subsection{Profile Fitting: King vs EFF}
Several different models have been used to describe star cluster density profiles. The more common choices include \citet[hereafter King]{King1962}, \citet[hereafter King66]{King1966}, \citet[hereafter Wilson]{Wilson1975}, \citet[hereafter EFF]{Elson1987} and \citet[hereafter Sersic]{Sersic1968}.

The empirical King model has been frequently used to describe the number surface density profiles of old globular clusters. King66 and Wilson models are single-mass, isotropic, modified isothermal sphere, defined by a phase-space distribution function. Wilson model differs from King66 in an additional term that produces clusters spatially more extended than King66 model. At small and intermediate radii, the two model profiles are quite similar. King66 and Wilson models have no analytic expressions for density or surface brightness. See \citet{McLaughlin2005} and \citet{McLaughlin2008} for a detailed comparison of both models. The other two models, EFF and Sersic, correspond with "power-law" models. They are defined to observationally describe more extended surface-density profiles. Like the Wilson model, Sersic model was originally designed for galaxy application.

The family of King model (King66 and King) are the most commonly used in star clusters studies, however \citet{McLaughlin2005} find that Wilson models fit clusters of any age, in any galaxy, as well as or better than King66 models. 
\citet{McLaughlin2008} conclude that analyses of structural correlations will not be significantly affected by using King66 models when Wilson profiles are better fits or vice versa. Based on this discussion we decide to fit a tidally truncated model and a more spatially extended model. For simplicity we chose the empirical King and EFF models to describe the surface brightness profiles of our sample.

The empirical King model \citep{King1962} is described as: 

\begin{equation}
\rho(R)= \rho_{0} \cdot \left\{ \frac{1}{\left[1+(R/R_{c})^{2}\right]^{\frac{1}{2}}} - \frac{1}{\left[1+(R_{t}/R_{c})^{2}\right]^{\frac{1}{2}}}\right\}^{2}
\label{King_model}
\end{equation}

where $\rho_{0}$ is the central surface number density of the cluster, $R_{c}$ is the core radius, and $R_{t}$ is the tidal radius. It is useful to define the concentration parameter as $c$ = log($R_{t}$/$R_{c}$). 

However, young LMC clusters do not appear to be tidally truncated, and seem to be more suitably fitted by power-law profiles. \citet{Elson1987} argue in favor of an unbound ``halo'' caused by the cluster expansion due to mass loss or violent relaxation, and proposed an empirical description known as the EFF model:

\begin{equation}
\mu(R) = \mu_{0} \cdot \left(1 + \frac{R^{2}}{a^{2}} \right)^{-\frac{\gamma}{2}}
\label{EFF_model}
\end{equation}

where $\mu_{0}$ is the central surface brightness, $a$ is the scale radius, and $\gamma$ is the power-law slope at large radii. The discrepancies between the two profiles are significant at large radii, being essentially indistinguishable in the inner regions.
The derived half-light radius, $R_{h}$, can be estimated based on the transformation equation in Larsen 2006. For the King model:    

\begin{equation}
R_{h} = 0.547 \cdot R_{c} \cdot (R_{t}/R_{c})^{0.486} 
\label{rh_King}
\end{equation}
and for the EFF profile:
\begin{equation}
R_{h} =  a \cdot \sqrt{(0.5)^{\frac{1}{(1-\frac{\gamma}{2})}} - 1}
\label{ep}
\end{equation}

Considering the presence of young clusters in our sample we decided to fit the cluster profiles with both types of models via weighted $\chi^{2}$ minimization. An additional constant term, $\phi_{bk}$, was added to the models to compensate for the background contamination. Although the clusters are quite well resolved, the effect of the point-spread function (PSF) in the core regions cannot be ignored. To improve the accuracy of the fitted parameters, the models were convolved with an appropriate PSF \citep[][for details]{SanRoman2009} before the minimization process. From the total sample of 161 clusters in Paper I, 33 objects were rejected due to low signal-to-noise ratios. Figure \ref{Fig5} shows the cluster surface brightness profiles with the best-fitting models where the error bars are estimations obtained by ELLIPSE from the isophotal intensities. Clusters with atypical profiles, Figure \ref{Fig6}, deviate from the analytical models in specific regions. These clusters have, in general, slightly larger $\chi^{2}$ values, relative to both models, than clusters with no anomalies. In fact the fit to some of these clusters are almost certainly not physically descriptive but even in these cases, the fits are sufficiently good to characterize the sizes of the clusters and their cores. The parameters from the best-fit King model are given in Table \ref{Table2}, with Table \ref{Table3} presenting the parameters from the best-fit EFF model.  We report the calibrated and extinction-corrected central surface brightness in the native band-pass of the data. Reddening values have been adopted from Paper I. When individual reddening is not available, we adopt a line-of-sight reddening value of E(V - I) = 0.06 \citep{Sarajedini2000}. The listed errors in the tables, are given by the $\chi^{2}$  minimization fitting process. Alternatively, the difference between the F606W and F814W fits can be used as a more realistic uncertainty.

\begin{figure*}
\begin{center}
\includegraphics[width=0.9\textwidth]{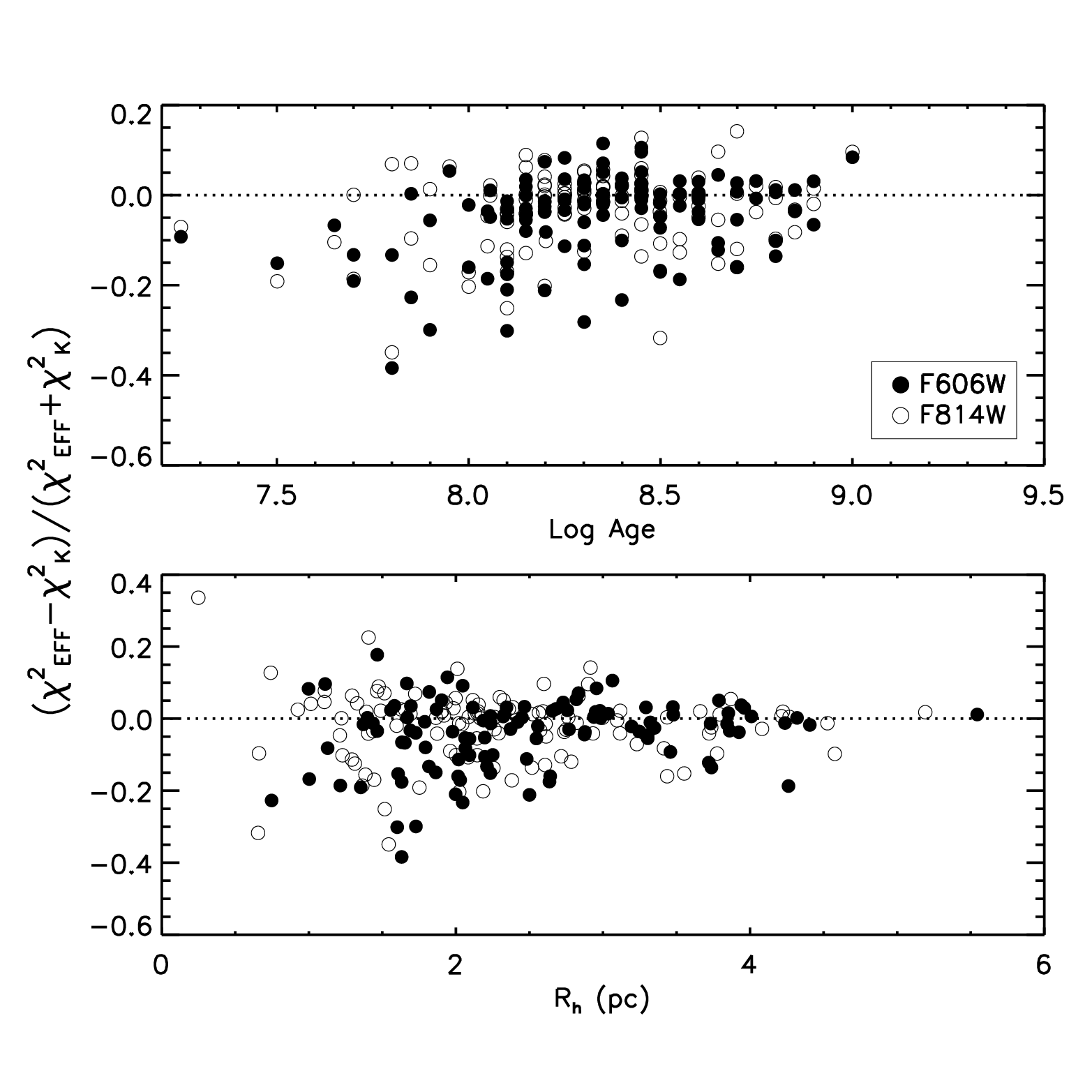}
\caption{Quality of the fit of EFF model relative to King model versus age and half-light radius for the two different filters.}
\label{Fig7}
\end{center}
\end{figure*}

Nine clusters, with available structural parameters, are common to \citet{Chandar1999}. Direct comparison between core radii gives us offsets of: $\Delta R_{c}$(CBF -- Us) = -- 0.81$\pm$ 0.3 pc. The profiles of those objects, except in two cases, have prominent irregularities in their profiles. \citet{Chandar1999} calculate core radii using a linear correlation with FWHM rather than direct model fits. The discrepancy between the two studies can be caused by the different adopted methods in addition to the peculiar profiles.

Following \citet{McLaughlin2005}, we define a statistic parameter that compares the $\chi^{2}$ of the best fit of both models for any object:

 \begin{equation}
\Delta = \frac{\chi^{2}_{EFF} - \chi^{2}_{King}}{\chi^{2}_{EFF} +  \chi^{2}_{King}}
\label{delta}
\end{equation}

If the parameter is zero, the two models fit the same cluster equally well. Positive values indicate a better fit of the King model, and negative values indicate EFF models are more suitable. Figure \ref{Fig7} shows this $\Delta$ parameter as a function of age and half-light radius. Estimated ages for our M33 sample were adopted from Paper I, where they were  obtained via color-magnitude diagram (CMD) fitting. Young clusters (Log age $<$ 8) are notably better fit by models with no radial truncation. In older clusters, small values of the statistic parameter, --0.2 $\leq \Delta \leq$ 0.2,  show no significant differences between the quality of the fittings. EFF models are also favored for smaller sizes as shown in the bottom panel of Figure \ref{Fig7}

\begin{figure*}
\begin{center}
\includegraphics[width=0.9\textwidth]{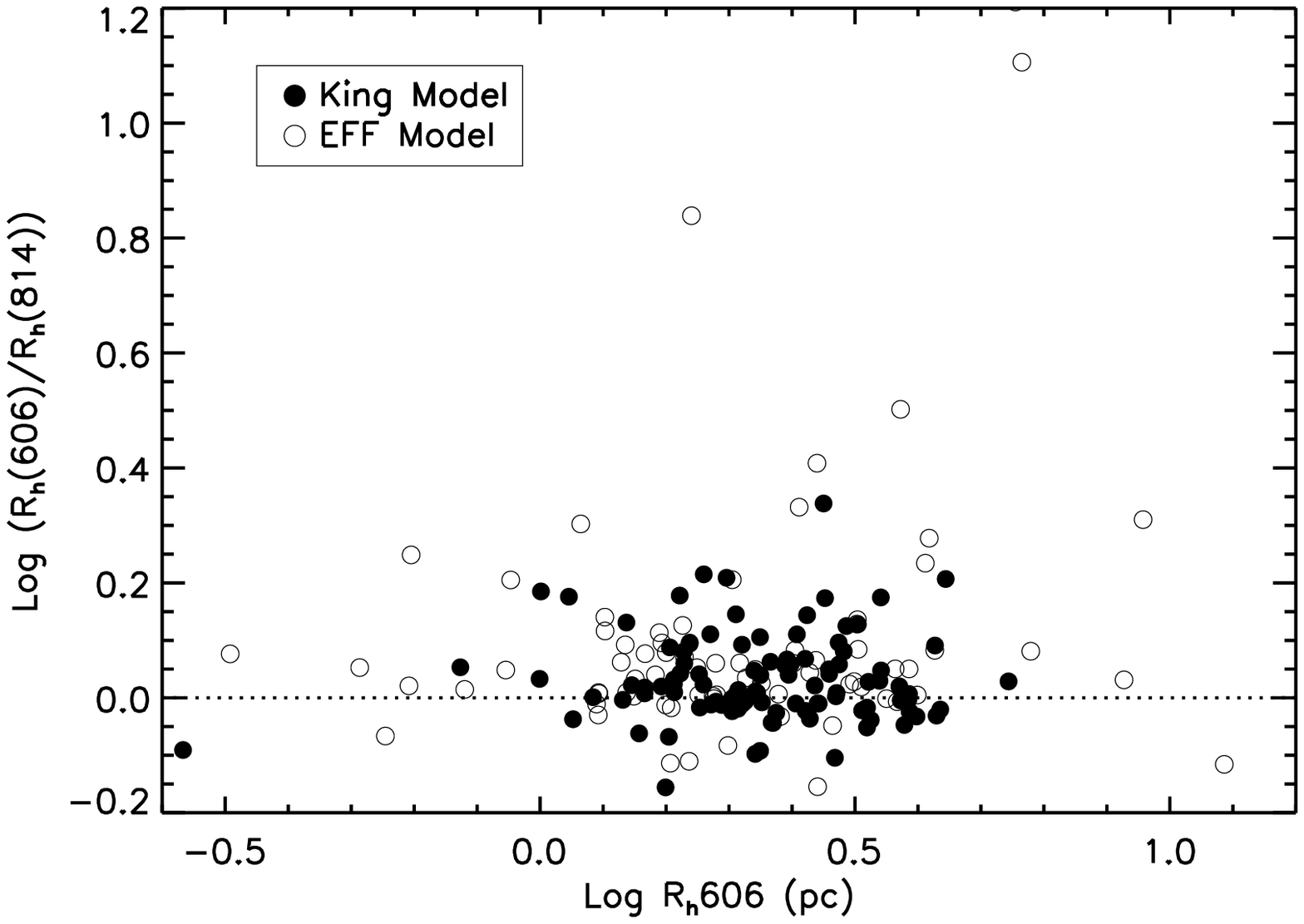}
\caption{Comparison of half-light radius observed in both F606W and F814W filters for the different fitted models.}
\label{Fig8}
\end{center}
\end{figure*}

Working with F606W and F814W provides us with two independent sets of measurements for each cluster. Systematic errors and color dependencies in the fits can be evaluated through direct comparison of both measurements. Figure \ref{Fig8} compares the parameters derived from model fits in F606W and F814W. The results show, in general, good agreement between bands, however specific clusters with significant disagreement between filters stand out. These differences in the half-light radii are more prominent for the EFF derived parameters. In the majority of the cases correspond with clusters with strong irregular profiles in both bands or anomalies in one filter and not in the other. In particular the derived structural parameters of clusters 3, 28, 33, 34, 70, 71, 95, 109, 116, 124, 159 are significantly in disagreement for the two different bands. We only consider the F606W model fits from now on because this filter is more sensitive to the underlying stellar population, and less sensitive to foreground contaminating stars. In addition, fits to clusters in other galaxies are typically performed in the V-band allowing us direct comparison without being concerned about possible color gradients. Since King models are primarily used to describe clusters in other galaxies, we adopt the derived King structural parameters as the basis for the subsequent analysis.

\section{Structural Parameters}

\subsection{Comparison with Other Galaxies}

The dynamical evolution of star clusters directly affects their structural properties. Therefore, comparing structural parameters in different galaxies is of special importance to understand star cluster evolution in a variety of environments. We have compared the newly derived M33 structural parameters with those for clusters in M31 \citep{Barmby2007}, the LMC and SMC \citep{McLaughlin2005}. M31 sample has been analyzed in an homogeneous way in the same filters F606W and F814W. Comparisons with LMC and SMC samples correspond with direct comparison with V-band photometric quantities. In addition, this study is based in number density converted to net mag arcsec$^{-2}$. Both studies have fitted a number of different models including King66.

While King model is described by the core radius, $R_{c}$, King66 is characterized by the scaled radius $R_{0}$. It is important to recognize that $R_{0}$ and $R_{c}$ are not strictly equivalent. For the King model, the core radius corresponds with the radius at which the density has fallen to just half the central density, $I_{King}(R_{c}) = 0.5 \cdot I_{0}$, while for King66 this relation is only approximate, $I_{King66}(R_{0}) \approx 0.5 \cdot I_{0}$. The theoretical models agree closely with the empirical family of curves for a reasonable range of values and fit the observations equally well \citep{King1966}. For this reason $R_{0}$ is often called the core radius. The connection between the theoretical and the observational parameters is straightforward to calculate and both studies of comparison provide, in addition to the basic King66 fit parameters, the derived structural parameters.  

\begin{figure*}
\begin{center}
\includegraphics[width=1\textwidth]{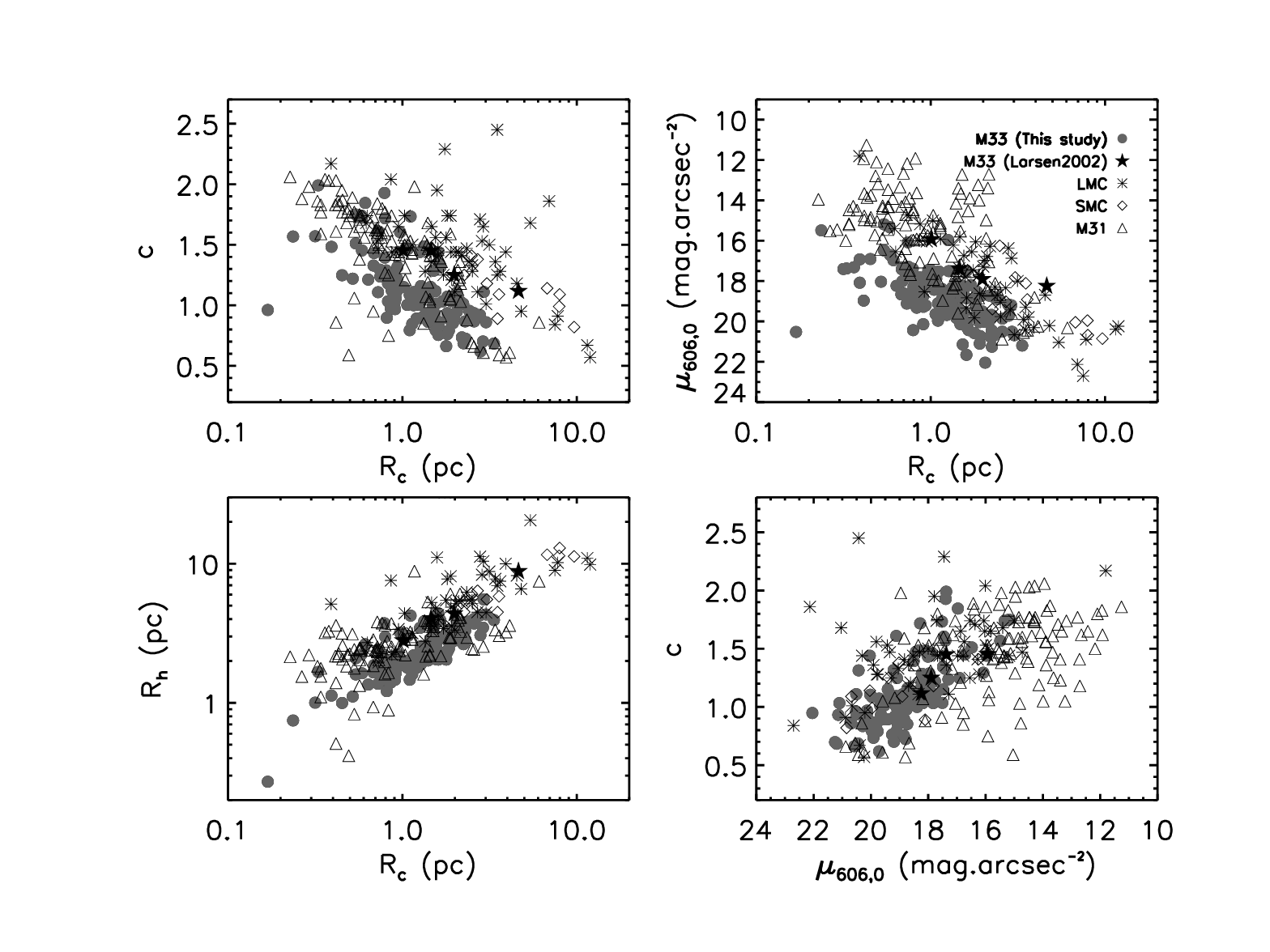}
\caption{Structural parameters of the present study as compared with star clusters in M31 \citep{Barmby2007}, LMC and SMC \citep{McLaughlin2005}. $R_{c}$ and $R_{h}$ are plotted in logarithmic scale for better comparison.}
\label{Fig9}
\end{center}
\end{figure*}

 Figure \ref{Fig9} shows correlations among different structural parameters. We have included the four old, halo M33 GCs analyzed in \citet{Larsen2002}. They appear to have similar structural properties than our sample except for the higher brightness. All the galaxies present similar trends although several differences stand out. M33 seems to possess clusters with small sizes, not showing objects with radii  between $R_{c}$= 3 -- 10 pc observed in other galaxies. With an average of $c$=1.12, the concentration in M33 clusters is smaller than the mean concentrations of clusters in the MW ($c$=1.41), M31 ($c$=1.46) and LMC ($c$=1.46). The top-right panel in Figure \ref{Fig9} shows  that the central surface brightness of M33 star clusters is fainter than the one in M31 and LMC star clusters, covering a smaller range and presenting a slightly offset position in the diagram. Direct comparison with V-band photometric quantities, in the case of LMC and SMC samples, could motivate the offset. In general more massive/luminous clusters tend to have brighter $\mu_{0}$. The brightest clusters in M33 are more than 1 mag fainter than the brightest clusters in MW, M31 and LMC \citep{SanRoman2010} so one would expect M33 star clusters to have fainter central surface brightness.

\begin{figure*}
\begin{center}
\includegraphics[width=1.\textwidth]{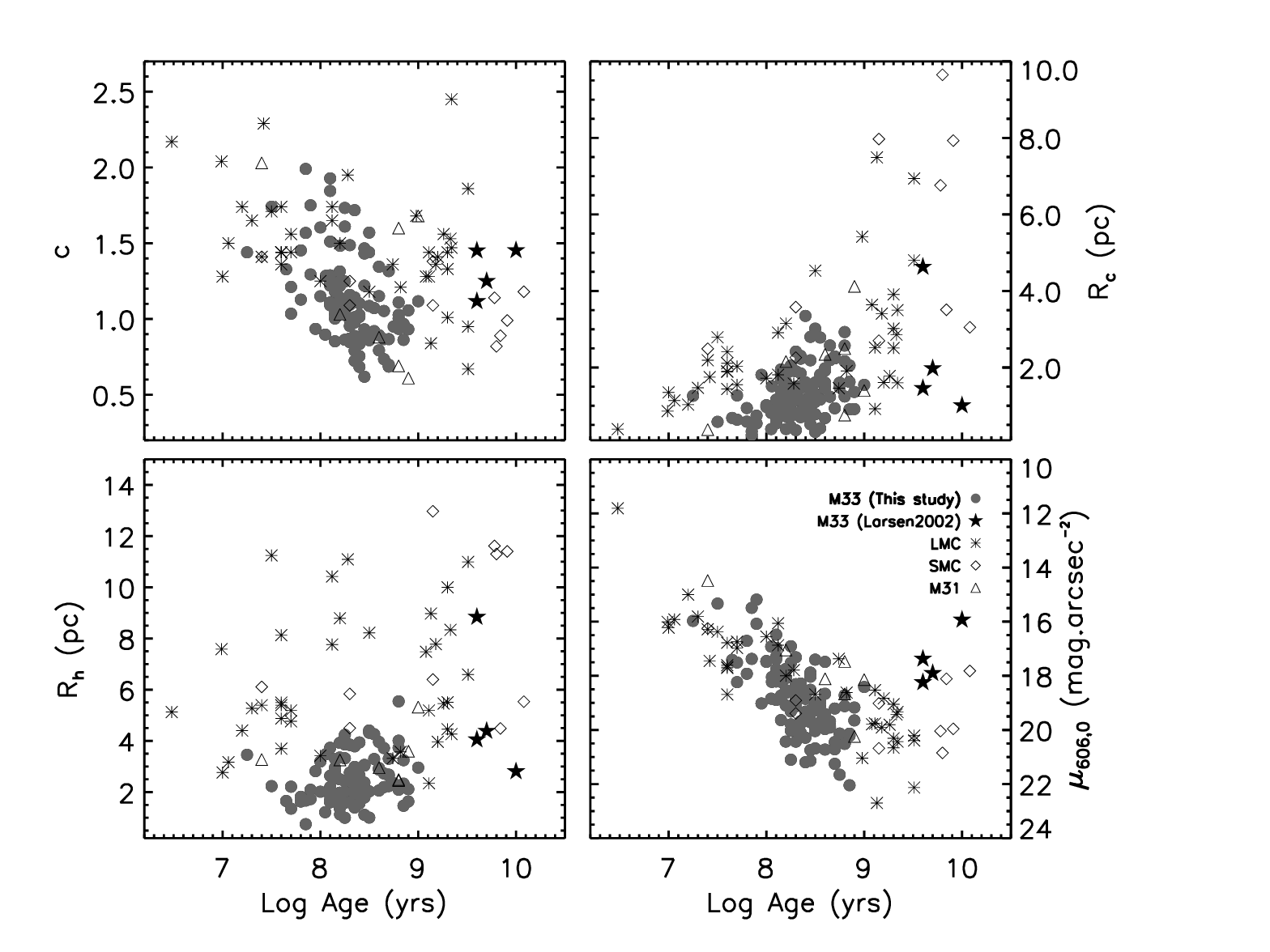}
\caption{Structural parameters of the present study versus age as compared with M31 \citep{Barmby2007}, LMC and SMC \citep{McLaughlin2005}}
\label{Fig10}
\end{center}
\end{figure*}

Structural properties for the M33 sample are shown as a function of estimated age in Figure \ref{Fig10}. When possible, we have compared our sample with the above studies. Ages for M31 clusters are adopted from \citet{Caldwell2009} for the common objects, while ages for LMC and SMC are provided in their previously mentioned studies. 

Comparison between these galaxies suggests similar trends. The age-radius relation among M33 clusters has been previously suggested \citep{Chandar1999} and also observed in different galaxies \citep[e.g.][]{Mackey2003a,Mackey2003b,Glatt2009}, suggesting that the age-radius trend has its origin in internal cluster processes, rather than external elements. \citet{Elson1989} identified, among LMC clusters, a clear trend that the spread in core radius is a increasing function of cluster age. This trend was confirmed later by \citet{Mackey2003a} using HST images. \citet{Elson1989} suggests mass loss from stellar evolution as the cause of cluster expansion. All clusters would have formed with small core radii and as the most massive stars evolve they would expand. In a more recent study, \citet{Mackey2008} use N-body simulations to specifically model the evolution of massive star clusters observed in LMC and SMC. They identify two physical processes which result in substantial and prolonged core expansion. The first process, mass-loss due to rapid stellar evolution in a primordially mass-segregated cluster, only occurs during the 100 -- 200 Myrs after the formation of the cluster. The second process, heating due to a retained population of stellar mass black holes, does not begin until several hundred Myr after the formation but result in core expansion for many Gyr. A combination effect of these two process result in a wide variety of evolutionary paths that cover the observed distribution of massive star clusters in LMC and SMC.  Several other mechanisms have been tested as possible explanations. \citet{Wilkinson2003} investigate the effect of a time-varying tidal field and variations in primordial binary fractions in N-body simulations. Assuming a point-mass galaxy the clusters show similar core radius evolution on both circular and elliptical orbits and therefore they conclude that the tidal field of the LMC does not influence the evolution of the clusters significantly. The presence of large numbers of primordial binaries in a cluster leads to core radius expansion; however the magnitude of the effect is insufficient to explain the observations in LMC clusters. 

The trend of older clusters having a larger range in core radii than the younger population is clearly visible in the top-right panel of Figure \ref{Fig10}. The youngest clusters in our sample possess compact cores of $R_{c}$ $\sim$ 0 -- 1 pc. At intermediate ages M33 clusters expand a range of 0 $\lesssim$ $R_{c}$ $\lesssim$ 4, slightly larger increase than among the LMC clusters in the same age period. A faster mass-loss rate could produce a rapid evolution of M33 clusters. Variations in the initial mass function (IMF) slope could be responsible for this effect. \citet{Gieles2008}, in an extensive survey of N-body simulations, show that the strength of the tidal field is the dominant factor determining the escape rate of stars. This result suggests that external elements can also produce an enhancement in the mass-loss rate and accelerate the evolution of the cluster. Strong environmental effect could also affect the evolution of M33 star clusters.

\subsection{Galactocentric Distribution}

Star clusters lie on the tidal field of their host galaxy and therefore are subject to tidal shocks. It is then reasonable to expect that their position from the center of the host galaxy can influence their dynamical evolution. Clusters closer to the center of the galaxy will be more dynamically evolved due to the effects of the tidal field. Structural parameters of MW globular clusters have been shown to be independent of galactocentric distance except for the half-light radius \citep[e.g.][]{vandenBergh1991, Djorgovski1994}. Similar results have been found in nearby galaxies \citep{Barmby2002,Barmby2007}. These studies show that star clusters at large galactocentric distances, on average, have larger diameters than those closer to the galactic center. While large clusters with small galactocentric distances could have been disrupted, there is no satisfactory explanation for the deficiency of small clusters at larger distances. For this reason, this 
correlation has been suggested to be an initial physical condition of cluster formation and a general property of star cluster systems \citep{vandenBergh1991}.

\begin{figure*}
\begin{center}
\includegraphics[width=0.9\textwidth]{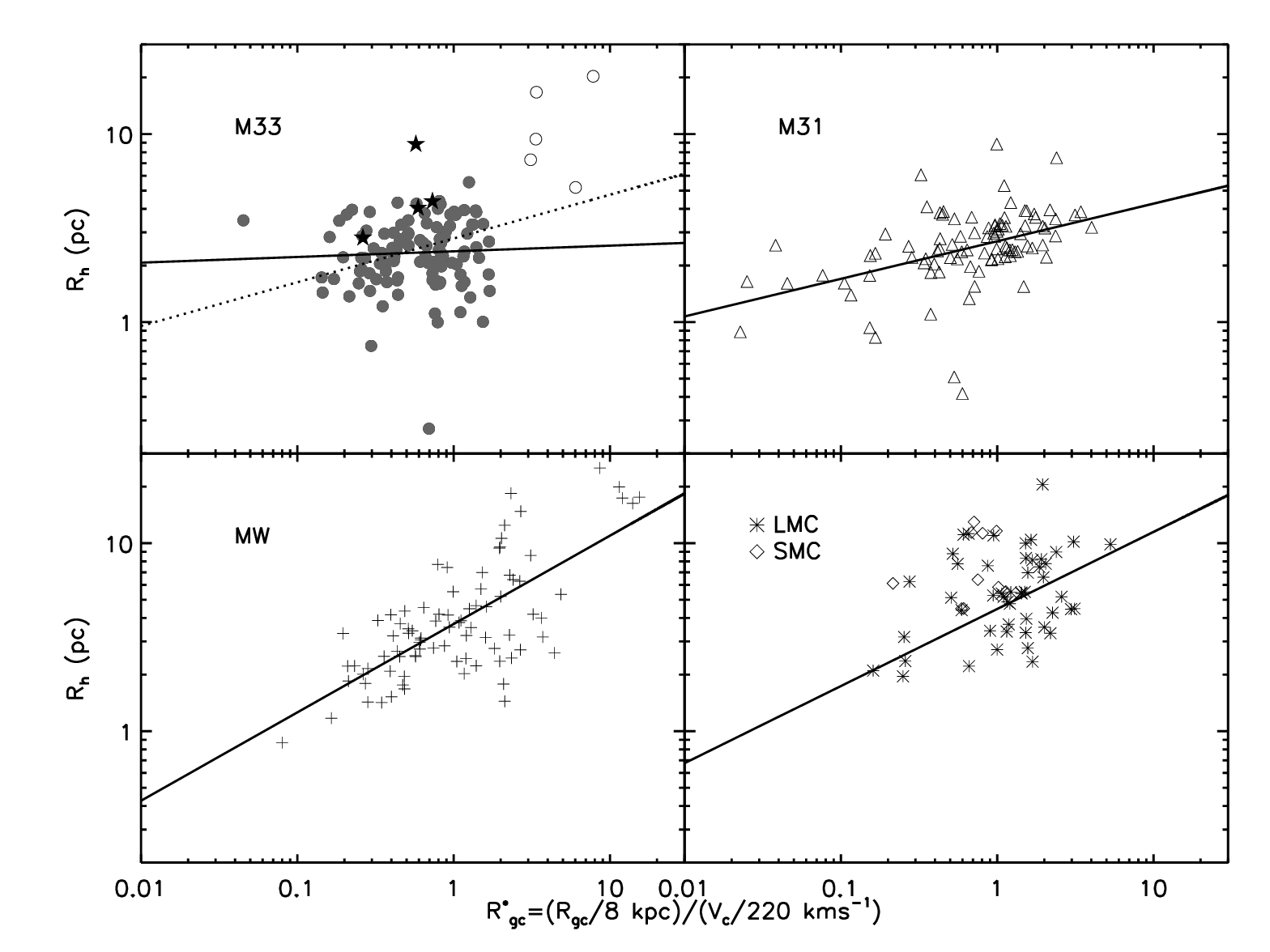}
\caption{Half-light radius as a function of renormalized galactocentric distance for clusters in our sample, M31, MW and MC. Solid lines represent the least-squares fits. In the top left panel, the solid line represents the least-squares fit including our sample (filled circles) and the four GCs in \citet{Larsen2002}(star symbols). Open circles in the same panel correspond with outskirt clusters in M33 \citep{Cockcroft2011}. The dashed line corresponds with the least-square fit including the outskirt cluster in M33.}
\label{Fig11}
\end{center}
\end{figure*}

Only projected distances are available for our sample. No correlation has been found between concentration and galactocentric distance, $R_{gc}$. Central surface brightness does not correlate either with $R_{gc}$. To explore the $R_{h}$ -- $R_{gc}$ relation, we follow \citet{Barmby2007}, and use the renormalized galactocentric distance as a better indicator of galactic potential than galactocentric distance itself:

\begin{equation}
R^{*}_{gc} = \frac{R_{gc} / (8 Kpc)}{V_{c} / (220 km s^{-1})}
\label{rh_King}
\end{equation}

where V$_{c}$ corresponds to the galaxy's circular velocity. We have used the following values of V$_{c}$ (in km s$^{-1}$): 220 for the MW, 100 for M33 \citep{Corbelli2000}, 230 for M31 \citep{Carignan2006}, 65 for the LMC \citep{vanderMarel2002} and 60 for the SMC \citep{Stanimirovic2004}. Figure \ref{Fig11} shows $R_{h}$ as a function of renormalized galactocentric distance, $R^{*}_{gc}$, for the different galaxies. The filled circles in the top left panel represent our sample of M33 clusters, and the star symbols correspond with the four globular clusters in \citet{Larsen2002}. The solid line corresponds with the least-squares fit of the combined sample. The solid lines for the rest of the galaxies have been taken from Table 10 \citep{Barmby2007} as the suggested star cluster fundamental plane fits. No clear correlation has been found in the M33 sample. Only 5 confirmed clusters have been detected in the outskirts of M33 ($R_{gc} >$ 10 Kpc) \citep{Stonkute2008,Huxor2009,Cockcroft2011}. \citet{Cockcroft2011} fit King models to these clusters and derived structural parameters. When these 5 clusters are included in the analysis, the data show a similar trend than in other galaxies (Top left panel of Figure \ref{Fig11}).
The five outer clusters are represented as open symbols, and the new least-squares fit corresponds with the dashed line.
While the trend in other galaxies is clearly driven by small inner clusters and also large clusters at larger radii, the correlation in M33 is just forced by very large clusters in the outer part of the galaxy. This could be a selection effect since at large distances, small clusters will be difficult to detect in ground-based images. We note that the star cluster systems in this analysis correspond with different age ranges, however as Figure \ref{Fig10} shows there is no correlation between $R_{h}$ with age. Therefore any differences between galaxies in Figure \ref{Fig11} are not due to secondary correlation with this parameter. In addition, \citet{Barmby2002,Barmby2007} find a mild correlation between $R_{h}$ and [Fe/H] in the MW, M31 and the MC. \citet{Brodie2006} discuss several possible explanations and argue that this correlation is a consequence of two effects, the $R_{h}$ -- $R_{gc}$ correlation and the predisposition of metal rich clusters to occupied smaller galactocentric radius.

High-resolution imaging has shown that the overall properties of GCs in different galaxies are remarkably similar. In fact, several studies suggest that GC structural and dynamical properties can be summarized as describing a fundamental plane analogous to the one for elliptical galaxies \citep{Djorgovski1995,Dubath1997, McLaughlin2000}. Dynamically and structurally, the four M33 clusters studied in \citet{Larsen2002}, appear virtually identical to MW and M31 globular clusters. Measurements of velocity dispersions for a large sample of M33 clusters would permit precise M/L ratios to determine with confidence if M33 clusters fall in the suggested fundamental plane for MW and M31 clusters.

\section{Summary}
We present ellipticities, position angles and structural parameters of M33 star clusters using \textit{HST/ACS} images in F606W and F814W bands. This study presents for the first time detailed morphological properties of a significant number of M33 star clusters. Surface brightness profiles have been fitted with King models and EFF models to determine structural properties such as as core and tidal radii, concentration and central surface brightness. On average, the clusters in M33 are more elliptical than those of the MW or M31 and more similar to clusters in the SMC. No correlation was found between the ellipticities and any of the other properties studied herein, including age or luminosity. The position angles of the clusters show a bimodality with a strong preferred orientation perpendicular to the position angle of M33 major axis.  A significant group of clusters shows irregularities such as bumps, dips or sharp edges in their surface brightness profiles. Young clusters are notably better fitted by models with no radial truncation (EFF models), while older clusters show no significant differences between King or EFF fits. M33 star clusters seem to have smaller sizes, smaller concentrations, and smaller central surface brightness as compared to clusters in the MW, M31, LMC and SMC. The sample presents an age-radius relation also detected in other galaxies. Structural parameters show no correlation with galactocentric distance although a trend with the half-light radius appears when combining our sample with M33 outer clusters. The overall analysis shows differences in the structural evolution between the M33 cluster system and cluster systems in nearby galaxies. These differences could have been caused by the strong differences in these various environments.

\section*{Acknowledgments}
We thank an anonymous referee for comments that greatly improved this paper. We are grateful for support from NASA through grant GO-10190 from the Space Telescope Science Institute, which is operated by the Association of Universities for Research in Astronomy, Inc., for NASA under contract NAS5-26555.

\newpage
\renewcommand{\arraystretch}{0.7}
\renewcommand{\thefootnote}{\alph{footnote}}

\small
\begin{longtable}{rcccc}
\caption{Ellipticities and Position Angle (PA).} \label{Table1} \\

\hline \hline \\
   Id$^{\mathsf{a}}$  &  Ellipticity &  Ellipticity  &   PA       & PA \\
                      &  (F606W)     &  (F814W )     &   (F606W)  & (F814W) \\\hline
   \\
\endfirsthead

\multicolumn{3}{c}{{\tablename} \thetable{} -- Continued} \\[0.5ex]
  \hline \hline \\
   Id$^{\mathsf{a}}$  &  Ellipticity &  Ellipticity  &   PA       & PA \\
                      &  (F606W)     &  (F814W )     &   (F606W)  & (F814W) \\\hline
   \\
\endhead


  \\
\endlastfoot


  1     &       0.23 $\pm$       0.01     &       0.29 $\pm$       0.02     &       63.4 $\pm$        2.6     &       64.2 $\pm$        2.2    \\
  2     &       0.25 $\pm$       0.04     &    ...                          &     ...                         &     ...                        \\
  3     &    ...                          &    ...                          &     ...                         &     ...                        \\
  4     &       0.14 $\pm$       0.03     &       0.18 $\pm$       0.04     &     ...                         &     ...                        \\
  5     &    ...                          &    ...                          &     ...                         &     ...                        \\
  6     &       0.17 $\pm$       0.02     &       0.17 $\pm$       0.03     &     ...                         &     ...                        \\
  7     &       0.23 $\pm$       0.02     &       0.25 $\pm$       0.01     &       64.5 $\pm$        2.0     &       62.7 $\pm$        1.5    \\
  8     &       0.05 $\pm$       0.01     &       0.14 $\pm$       0.02     &     ...                         &     ...                        \\
  9     &    ...                          &    ...                          &     ...                         &     ...                        \\
 10     &       0.17 $\pm$       0.03     &    ...                          &     ...                         &     ...                        \\
 11     &       0.14 $\pm$       0.02     &       0.20 $\pm$       0.02     &     ...                         &     ...                        \\
 12     &       0.30 $\pm$       0.04     &       0.33 $\pm$       0.03     &     ...                         &     ...                        \\
 13     &       0.16 $\pm$       0.03     &    ...                          &     ...                         &     ...                        \\
 14     &       0.22 $\pm$       0.03     &       0.22 $\pm$       0.02     &       63.1 $\pm$        4.1     &       62.4 $\pm$        5.4    \\
 15     &       0.43 $\pm$       0.03     &       0.38 $\pm$       0.02     &      -40.9 $\pm$        4.3     &      -52.7 $\pm$        5.7    \\
 16     &       0.21 $\pm$       0.02     &       0.22 $\pm$       0.03     &     ...                         &     ...                        \\
 17     &       0.15 $\pm$       0.02     &       0.18 $\pm$       0.03     &     ...                         &     ...                        \\
 18     &       0.29 $\pm$       0.04     &    ...                          &     ...                         &     ...                        \\
 19     &       0.13 $\pm$       0.02     &       0.10 $\pm$       0.01     &      -46.3 $\pm$        3.9     &      -45.6 $\pm$        3.8    \\
 20     &       0.19 $\pm$       0.04     &       0.17 $\pm$       0.03     &     ...                         &     ...                        \\
 21     &       0.36 $\pm$       0.02     &    ...                          &     ...                         &     ...                        \\
 22     &       0.06 $\pm$       0.01     &       0.09 $\pm$       0.02     &     ...                         &     ...                        \\
 23     &       0.28 $\pm$       0.03     &    ...                          &     ...                         &     ...                        \\
 24     &       0.24 $\pm$       0.02     &       0.15 $\pm$       0.02     &     ...                         &     ...                        \\
 25     &       0.19 $\pm$       0.02     &       0.25 $\pm$       0.03     &       41.7 $\pm$        4.2     &       37.8 $\pm$        5.6    \\
 26     &       0.26 $\pm$       0.03     &       0.29 $\pm$       0.02     &      -62.0 $\pm$        4.3     &      -58.2 $\pm$        4.4    \\
 27     &       0.12 $\pm$       0.03     &       0.13 $\pm$       0.02     &     ...                         &     ...                        \\
 28     &       0.14 $\pm$       0.00     &    ...                          &      -70.0 $\pm$        0.0     &     ...                        \\
 29     &       0.25 $\pm$       0.05     &       0.33 $\pm$       0.04     &     ...                         &     ...                        \\
 30     &       0.23 $\pm$       0.01     &       0.23 $\pm$       0.01     &      -69.4 $\pm$        3.2     &      -66.4 $\pm$        3.6    \\
 31     &       0.24 $\pm$       0.03     &       0.26 $\pm$       0.02     &     ...                         &     ...                        \\
 32     &       0.24 $\pm$       0.02     &       0.26 $\pm$       0.02     &     ...                         &     ...                        \\
 33     &       0.09 $\pm$       0.01     &    ...                          &      -30.0 $\pm$        5.1     &     ...                        \\
 34     &    ...                          &    ...                          &     ...                         &     ...                        \\
 35     &       0.27 $\pm$       0.03     &       0.37 $\pm$       0.02     &     ...                         &     ...                        \\
 36     &       0.14 $\pm$       0.00     &       0.22 $\pm$       0.00     &      -70.0 $\pm$        0.0     &      -75.1 $\pm$        0.0    \\
 37     &       0.23 $\pm$       0.02     &       0.18 $\pm$       0.03     &       67.5 $\pm$        1.2     &      -28.0 $\pm$       25.2    \\
 38     &       0.12 $\pm$       0.00     &       0.17 $\pm$       0.03     &      -58.3 $\pm$        0.0     &      -23.6 $\pm$       17.4    \\
 39     &       0.45 $\pm$       0.02     &       0.40 $\pm$       0.01     &       38.6 $\pm$        3.2     &       30.9 $\pm$        0.4    \\
 40     &       0.21 $\pm$       0.02     &       0.19 $\pm$       0.03     &     ...                         &     ...                        \\

\hline\hline
\multicolumn{5}{l}{Note. -- Table \ref{Table1} is published in its entirety in the electronic edition of the Journal.} \\ 
\multicolumn{5}{l}{$^{\mathsf{a}}$ Identification number in Paper I} \\
\end{longtable}
\normalsize
\renewcommand{\arraystretch}{1.0}


\newpage
\begin{landscape}
\renewcommand{\arraystretch}{0.7}
\renewcommand{\thefootnote}{\alph{footnote}}

\small
\begin{longtable}{rccccccccccc}
\caption{Structural Parameters from the King Profile Fit.} \label{Table2} \\

\hline \hline \\
   Id$^{\mathsf{a}}$ &  Filter &  $\mu_{0}$          & R$_{c}$  &  R$_{c}$     & R$_{t}$  &  R$_{t}$     &   c  & R$_{h}$   & R$_{h}$  &$\phi_{bk}$ \\
                   &         &  (mag arsec$^{-2}$)  & (arcsec) &  (pc)   &   (arcsec) &  (pc)  &        & (arcsec)  & (pc)    & (L$_{\odot}$pc$^{-2}$) \\ \hline 
   \\
\endfirsthead
\multicolumn{3}{c}{{\tablename} \thetable{} -- Continued} \\[0.5ex]
  \hline \hline \\
   Id$^{\mathsf{a}}$ &  Filter &  $\mu_{0}$          & R$_{c}$  &  R$_{c}$     & R$_{t}$  &  R$_{t}$     &   c  & R$_{h}$   & R$_{h}$  &$\phi_{bk}$ \\
                   &         &  (mag arsec$^{-2}$)  & (arcsec) &  (pc)   &   (arcsec) &  (pc)  &        & (arcsec)  & (pc)    & (L$_{\odot}$pc$^{-2}$) \\ \hline 
   \\
\endhead


  \\
\endlastfoot


    1 &    F606W &    19.17 $\pm$  0.02    &   0.22 $\pm$  0.00    &   0.92 $\pm$  0.02    &       2.47 $\pm$      0.09    &      10.43 $\pm$      0.36    &       1.06 $\pm$  0.02    &     0.39 $\pm$  0.01    &       1.63 $\pm$      0.03    &    34.96 $\pm$  0.25     \\
      &    F814W &    18.55 $\pm$  0.02    &   0.21 $\pm$  0.00    &   0.90 $\pm$  0.02    &       2.41 $\pm$      0.08    &      10.16 $\pm$      0.33    &       1.05 $\pm$  0.02    &     0.38 $\pm$  0.01    &       1.60 $\pm$      0.03    &    39.82 $\pm$  0.28     \\
    2 &    F606W &    18.97 $\pm$  0.02    &   0.10 $\pm$  0.00    &   0.41 $\pm$  0.01    &      ...                      &      ...                      &      ...                  &    ...                  &      ...                      &    37.47 $\pm$  0.28     \\
      &    F814W &    18.89 $\pm$  0.01    &   0.14 $\pm$  0.00    &   0.57 $\pm$  0.01    &      ...                      &      ...                      &      ...                  &    ...                  &      ...                      &    42.16 $\pm$  0.26     \\
    3 &    F606W &    21.19 $\pm$  0.06    &   0.79 $\pm$  0.04    &   3.35 $\pm$  0.16    &       3.84 $\pm$      0.19    &      16.19 $\pm$      0.79    &       0.68 $\pm$  0.03    &     0.93 $\pm$  0.03    &       3.94 $\pm$      0.13    &    36.90 $\pm$  0.32     \\
      &    F814W &    20.78 $\pm$  0.10    &   0.32 $\pm$  0.02    &   1.34 $\pm$  0.08    &      ...                      &      ...                      &      ...                  &    ...                  &      ...                      &    42.50 $\pm$  0.38     \\
    4 &    F606W &    18.08 $\pm$  0.02    &   0.09 $\pm$  0.00    &   0.39 $\pm$  0.01    &       2.83 $\pm$      0.22    &      11.92 $\pm$      0.92    &       1.48 $\pm$  0.04    &     0.27 $\pm$  0.01    &       1.13 $\pm$      0.04    &    36.71 $\pm$  0.20     \\
      &    F814W &    18.15 $\pm$  0.01    &   0.10 $\pm$  0.00    &   0.44 $\pm$  0.01    &       3.00 $\pm$      0.37    &      12.65 $\pm$      1.55    &       1.46 $\pm$  0.05    &     0.29 $\pm$  0.02    &       1.23 $\pm$      0.07    &    43.17 $\pm$  0.26     \\
    5 &    F606W &    20.43 $\pm$  0.03    &   0.19 $\pm$  0.00    &   0.79 $\pm$  0.02    &      ...                      &      ...                      &      ...                  &    ...                  &      ...                      &    42.60 $\pm$  0.36     \\
      &    F814W &    20.43 $\pm$  0.08    &   0.22 $\pm$  0.01    &   0.94 $\pm$  0.05    &       4.00 $\pm$      1.65    &      16.86 $\pm$      6.95    &       1.25 $\pm$  0.18    &     0.50 $\pm$  0.10    &       2.09 $\pm$      0.42    &    50.11 $\pm$  0.39     \\
    6 &    F606W &    19.21 $\pm$  0.02    &   0.22 $\pm$  0.00    &   0.94 $\pm$  0.02    &       4.60 $\pm$      0.67    &      19.41 $\pm$      2.84    &       1.32 $\pm$  0.06    &     0.53 $\pm$  0.04    &       2.23 $\pm$      0.16    &    40.15 $\pm$  0.29     \\
      &    F814W &    18.83 $\pm$  0.03    &   0.22 $\pm$  0.01    &   0.91 $\pm$  0.03    &       7.37 $\pm$      2.64    &      31.09 $\pm$     11.14    &       1.53 $\pm$  0.16    &     0.65 $\pm$  0.11    &       2.76 $\pm$      0.48    &    45.54 $\pm$  0.39     \\
    7 &    F606W &    21.66 $\pm$  0.07    &   0.38 $\pm$  0.02    &   1.60 $\pm$  0.08    &      ...                      &      ...                      &      ...                  &    ...                  &      ...                      &    25.56 $\pm$  0.22     \\
      &    F814W &    21.72 $\pm$  0.06    &   0.38 $\pm$  0.01    &   1.58 $\pm$  0.06    &      ...                      &      ...                      &      ...                  &    ...                  &      ...                      &    29.47 $\pm$  0.17     \\
    8 &    F606W &    19.66 $\pm$  0.03    &   0.29 $\pm$  0.01    &   1.21 $\pm$  0.03    &       2.22 $\pm$      0.08    &       9.38 $\pm$      0.35    &       0.89 $\pm$  0.02    &     0.42 $\pm$  0.01    &       1.79 $\pm$      0.04    &    39.05 $\pm$  0.38     \\
      &    F814W &    19.17 $\pm$  0.03    &   0.27 $\pm$  0.01    &   1.13 $\pm$  0.03    &       1.95 $\pm$      0.06    &       8.24 $\pm$      0.27    &       0.86 $\pm$  0.02    &     0.39 $\pm$  0.01    &       1.63 $\pm$      0.03    &    43.73 $\pm$  0.43     \\
   10 &    F606W &    20.03 $\pm$  0.04    &   0.49 $\pm$  0.02    &   2.05 $\pm$  0.07    &       6.19 $\pm$      0.79    &      26.10 $\pm$      3.33    &       1.11 $\pm$  0.06    &     0.92 $\pm$  0.06    &       3.86 $\pm$      0.25    &    41.73 $\pm$  0.25     \\
      &    F814W &    20.22 $\pm$  0.05    &   0.65 $\pm$  0.02    &   2.73 $\pm$  0.10    &       5.11 $\pm$      0.51    &      21.57 $\pm$      2.15    &       0.90 $\pm$  0.05    &     0.97 $\pm$  0.05    &       4.08 $\pm$      0.21    &    46.59 $\pm$  0.34     \\
   11 &    F606W &    19.86 $\pm$  0.03    &   0.61 $\pm$  0.01    &   2.58 $\pm$  0.06    &       4.50 $\pm$      0.22    &      18.96 $\pm$      0.91    &       0.87 $\pm$  0.02    &     0.88 $\pm$  0.02    &       3.72 $\pm$      0.10    &    44.13 $\pm$  0.39     \\
      &    F814W &    19.17 $\pm$  0.03    &   0.59 $\pm$  0.02    &   2.49 $\pm$  0.07    &       4.24 $\pm$      0.20    &      17.88 $\pm$      0.84    &       0.86 $\pm$  0.02    &     0.84 $\pm$  0.02    &       3.55 $\pm$      0.09    &    49.36 $\pm$  0.57     \\
   12 &    F606W &    20.14 $\pm$  0.02    &   0.22 $\pm$  0.00    &   0.91 $\pm$  0.02    &       2.00 $\pm$      0.06    &       8.42 $\pm$      0.27    &       0.97 $\pm$  0.02    &     0.35 $\pm$  0.01    &       1.47 $\pm$      0.03    &    25.01 $\pm$  0.15     \\
      &    F814W &    19.70 $\pm$  0.02    &   0.22 $\pm$  0.00    &   0.95 $\pm$  0.02    &       1.83 $\pm$      0.05    &       7.74 $\pm$      0.22    &       0.91 $\pm$  0.02    &     0.34 $\pm$  0.01    &       1.44 $\pm$      0.03    &    28.48 $\pm$  0.17     \\
   13 &    F606W &    18.77 $\pm$  0.03    &   0.21 $\pm$  0.01    &   0.90 $\pm$  0.03    &       2.28 $\pm$      0.08    &       9.60 $\pm$      0.32    &       1.03 $\pm$  0.02    &     0.37 $\pm$  0.01    &       1.56 $\pm$      0.03    &    40.74 $\pm$  0.38     \\
      &    F814W &    18.38 $\pm$  0.02    &   0.22 $\pm$  0.00    &   0.93 $\pm$  0.02    &       2.02 $\pm$      0.07    &       8.50 $\pm$      0.29    &       0.96 $\pm$  0.02    &     0.35 $\pm$  0.01    &       1.49 $\pm$      0.03    &    46.60 $\pm$  0.50     \\
   14 &    F606W &    18.80 $\pm$  0.02    &   0.44 $\pm$  0.01    &   1.84 $\pm$  0.04    &       5.03 $\pm$      0.22    &      21.20 $\pm$      0.94    &       1.06 $\pm$  0.02    &     0.78 $\pm$  0.02    &       3.30 $\pm$      0.08    &    24.96 $\pm$  0.29     \\
      &    F814W &    19.17 $\pm$  0.03    &   0.48 $\pm$  0.01    &   2.01 $\pm$  0.05    &       5.85 $\pm$      0.43    &      24.67 $\pm$      1.81    &       1.09 $\pm$  0.03    &     0.88 $\pm$  0.03    &       3.72 $\pm$      0.14    &   183.60 $\pm$  0.29     \\
   17 &    F606W &    20.71 $\pm$  0.03    &   0.54 $\pm$  0.02    &   2.29 $\pm$  0.08    &       3.99 $\pm$      0.22    &      16.85 $\pm$      0.94    &       0.87 $\pm$  0.03    &     0.78 $\pm$  0.03    &       3.30 $\pm$      0.11    &    23.81 $\pm$  0.22     \\
      &    F814W &    ...                  &   0.70 $\pm$  0.03    &   2.95 $\pm$  0.12    &       3.31 $\pm$      0.08    &      13.98 $\pm$      0.34    &       0.68 $\pm$  0.02    &     0.81 $\pm$  0.02    &       3.44 $\pm$      0.08    &   181.31 $\pm$  0.30     \\
   18 &    F606W &    18.23 $\pm$  0.03    &   0.15 $\pm$  0.01    &   0.64 $\pm$  0.03    &       2.46 $\pm$      0.10    &      10.39 $\pm$      0.42    &       1.21 $\pm$  0.03    &     0.32 $\pm$  0.01    &       1.35 $\pm$      0.04    &    27.36 $\pm$  0.41     \\
      &    F814W &    18.68 $\pm$  0.03    &   0.18 $\pm$  0.01    &   0.76 $\pm$  0.03    &       2.08 $\pm$      0.06    &       8.77 $\pm$      0.27    &       1.06 $\pm$  0.02    &     0.32 $\pm$  0.01    &       1.36 $\pm$      0.04    &   185.03 $\pm$  0.40     \\
   19 &    F606W &    18.51 $\pm$  0.01    &   0.19 $\pm$  0.00    &   0.79 $\pm$  0.02    &       4.66 $\pm$      0.45    &      19.63 $\pm$      1.91    &       1.39 $\pm$  0.04    &     0.49 $\pm$  0.02    &       2.06 $\pm$      0.10    &    45.80 $\pm$  0.28     \\
      &    F814W &    18.03 $\pm$  0.02    &   0.20 $\pm$  0.00    &   0.86 $\pm$  0.02    &       4.03 $\pm$      0.33    &      16.98 $\pm$      1.38    &       1.30 $\pm$  0.04    &     0.47 $\pm$  0.02    &       2.00 $\pm$      0.08    &    51.76 $\pm$  0.36     \\
   22 &    F606W &    18.72 $\pm$  0.01    &   0.51 $\pm$  0.00    &   2.16 $\pm$  0.02    &       5.46 $\pm$      0.17    &      23.05 $\pm$      0.70    &       1.03 $\pm$  0.01    &     0.89 $\pm$  0.01    &       3.74 $\pm$      0.06    &    45.29 $\pm$  0.28     \\
      &    F814W &    17.97 $\pm$  0.01    &   0.54 $\pm$  0.01    &   2.27 $\pm$  0.03    &       5.31 $\pm$      0.15    &      22.39 $\pm$      0.63    &       0.99 $\pm$  0.01    &     0.89 $\pm$  0.01    &       3.78 $\pm$      0.06    &    52.70 $\pm$  0.39     \\

\hline\hline
\multicolumn{10}{l}{Note. -- Table \ref{Table2} is published in its entirety in the electronic edition of the Journal.} \\ 
\multicolumn{10}{l}{$^{\mathsf{a}}$ Identification number in Paper I} \\
\end{longtable}
\normalsize
\renewcommand{\arraystretch}{1.0}
\end{landscape}


\newpage
\begin{landscape}
\renewcommand{\arraystretch}{0.7}
\renewcommand{\thefootnote}{\alph{footnote}}

\small
\begin{longtable}{rccccccccc}
\caption{Structural Parameters from the EFF Profile Fit.} \label{Table3} \\

\hline \hline \\
   Id$^{\mathsf{a}}$ &  Filter &  $\mu_{0}$          & a$_{c}$  &  a$_{c}$ &  $\gamma$ & R$_{h}$   & R$_{h}$  &$\phi_{bk}$\\
                   &         &  (mag arsec$^{-2}$)  & (arcsec) &  (pc)   &           & (arcsec)  & (pc)    & (L$_{\odot}$pc$^{-2}$)\\ \hline
   \\
\endfirsthead
\multicolumn{3}{c}{{\tablename} \thetable{} -- Continued} \\[0.5ex]
  \hline \hline \\
  Id$^{\mathsf{a}}$ &  Filter &  $\mu_{0}$ & a$_{c}$ &  a$_{c}$ &  $\gamma$ & R$_{h}$  & R$_{h}$  &$\phi_{bk}$ \\
                   &         &  (mag arsec$^{-2}$)  & (arcsec) &  (pc)   &           & (arcsec)  & (pc)    & (L$_{\odot}$pc$^{-2}$)\\ \hline
   \\
\endhead


  \\
\endlastfoot

    1    &    F606W    &    19.43 $\pm$  0.02    &   0.39 $\pm$  0.03    &   1.66 $\pm$  0.11    &   4.56 $\pm$  0.38    &     0.33 $\pm$  0.04    &     1.00 $\pm$  0.00    &    36.76 $\pm$  0.22       \\
         &    F814W    &    18.80 $\pm$  0.02    &   0.35 $\pm$  0.02    &   1.48 $\pm$  0.09    &   4.16 $\pm$  0.30    &     0.33 $\pm$  0.04    &     1.00 $\pm$  0.00    &    41.90 $\pm$  0.25      \\ 
    2    &    F606W    &    18.93 $\pm$  0.02    &   0.08 $\pm$  0.00    &   0.32 $\pm$  0.02    &   1.65 $\pm$  0.06    &     0.08 $\pm$  0.00    &     0.00 $\pm$  0.00    &    36.74 $\pm$  0.29       \\
         &    F814W    &    18.78 $\pm$  0.02    &   0.07 $\pm$  0.00    &   0.29 $\pm$  0.02    &   1.29 $\pm$  0.04    &     0.06 $\pm$  0.00    &     0.00 $\pm$  0.00    &    39.99 $\pm$  0.40      \\ 
    3    &    F606W    &    22.05 $\pm$  0.05    &   1.19 $\pm$  0.23    &   5.00 $\pm$  0.98    &   5.13 $\pm$  1.47    &     0.89 $\pm$  0.29    &     3.00 $\pm$  1.00    &    37.89 $\pm$  0.26       \\
         &    F814W    &    20.85 $\pm$  0.04    &   0.28 $\pm$  0.03    &   1.18 $\pm$  0.14    &   1.81 $\pm$  0.21    &     0.28 $\pm$  0.03    &     1.00 $\pm$  0.00    &    41.91 $\pm$  0.42      \\ 
    4    &    F606W    &    18.20 $\pm$  0.02    &   0.13 $\pm$  0.01    &   0.54 $\pm$  0.03    &   2.76 $\pm$  0.10    &     0.29 $\pm$  0.05    &     1.00 $\pm$  0.00    &    37.22 $\pm$  0.19       \\
         &    F814W    &    18.27 $\pm$  0.02    &   0.14 $\pm$  0.01    &   0.61 $\pm$  0.03    &   2.79 $\pm$  0.12    &     0.31 $\pm$  0.06    &     1.00 $\pm$  0.00    &    43.67 $\pm$  0.26      \\ 
    5    &    F606W    &    20.38 $\pm$  0.03    &   0.12 $\pm$  0.01    &   0.51 $\pm$  0.04    &   1.31 $\pm$  0.08    &     0.11 $\pm$  0.01    &     0.00 $\pm$  0.00    &    41.20 $\pm$  0.40       \\
         &    F814W    &    20.59 $\pm$  0.04    &   0.21 $\pm$  0.03    &   0.88 $\pm$  0.10    &   2.06 $\pm$  0.25    &    ...                  &    ...                  &    49.67 $\pm$  0.50      \\ 
    6    &    F606W    &    19.33 $\pm$  0.02    &   0.25 $\pm$  0.01    &   1.08 $\pm$  0.05    &   2.50 $\pm$  0.13    &     0.98 $\pm$  0.41    &     4.00 $\pm$  1.00    &    40.16 $\pm$  0.32       \\
         &    F814W    &    18.90 $\pm$  0.03    &   0.23 $\pm$  0.01    &   0.97 $\pm$  0.06    &   2.23 $\pm$  0.12    &    ...                  &    ...                  &    45.23 $\pm$  0.41      \\ 
    7    &    F606W    &    21.63 $\pm$  0.05    &   0.32 $\pm$  0.03    &   1.37 $\pm$  0.11    &   1.77 $\pm$  0.12    &     0.32 $\pm$  0.03    &     1.00 $\pm$  0.00    &    25.28 $\pm$  0.19       \\
         &    F814W    &    21.62 $\pm$  0.05    &   0.27 $\pm$  0.02    &   1.15 $\pm$  0.10    &   1.47 $\pm$  0.10    &     0.26 $\pm$  0.02    &     1.00 $\pm$  0.00    &    28.93 $\pm$  0.22      \\ 
    8    &    F606W    &    20.02 $\pm$  0.02    &   0.56 $\pm$  0.06    &   2.38 $\pm$  0.27    &   6.04 $\pm$  0.98    &     0.36 $\pm$  0.06    &     1.00 $\pm$  0.00    &    41.32 $\pm$  0.26       \\
         &    F814W    &    19.56 $\pm$  0.03    &   0.55 $\pm$  0.07    &   2.33 $\pm$  0.31    &   6.55 $\pm$  1.28    &     0.33 $\pm$  0.06    &     1.00 $\pm$  0.00    &    46.55 $\pm$  0.30      \\ 
   10    &    F606W    &    20.25 $\pm$  0.02    &   0.69 $\pm$  0.05    &   2.89 $\pm$  0.23    &   3.36 $\pm$  0.30    &     0.91 $\pm$  0.19    &     3.00 $\pm$  0.00    &    42.04 $\pm$  0.27       \\
         &    F814W    &    20.66 $\pm$  0.01    &   1.15 $\pm$  0.17    &   4.84 $\pm$  0.73    &   5.41 $\pm$  1.23    &     0.81 $\pm$  0.20    &     3.00 $\pm$  0.00    &    47.47 $\pm$  0.35      \\ 
   11    &    F606W    &    20.26 $\pm$  0.01    &   1.34 $\pm$  0.13    &   5.64 $\pm$  0.53    &   7.24 $\pm$  1.05    &     0.74 $\pm$  0.09    &     3.00 $\pm$  0.00    &    46.08 $\pm$  0.31       \\
         &    F814W    &    19.58 $\pm$  0.01    &   1.39 $\pm$  0.15    &   5.86 $\pm$  0.63    &   8.19 $\pm$  1.39    &     0.70 $\pm$  0.09    &     2.00 $\pm$  0.00    &    52.48 $\pm$  0.44      \\ 
   12    &    F606W    &    20.45 $\pm$  0.01    &   0.49 $\pm$  0.06    &   2.08 $\pm$  0.25    &   6.54 $\pm$  1.08    &     0.29 $\pm$  0.05    &     1.00 $\pm$  0.00    &    26.12 $\pm$  0.10       \\
         &    F814W    &    20.07 $\pm$  0.01    &   0.63 $\pm$  0.11    &   2.67 $\pm$  0.45    &   9.34 $\pm$  2.43    &     0.29 $\pm$  0.06    &     1.00 $\pm$  0.00    &    29.79 $\pm$  0.12      \\ 
   13    &    F606W    &    19.02 $\pm$  0.03    &   0.33 $\pm$  0.02    &   1.41 $\pm$  0.10    &   3.99 $\pm$  0.32    &     0.34 $\pm$  0.04    &     1.00 $\pm$  0.00    &    43.26 $\pm$  0.30       \\
         &    F814W    &    18.66 $\pm$  0.02    &   0.37 $\pm$  0.03    &   1.57 $\pm$  0.13    &   4.60 $\pm$  0.50    &     0.31 $\pm$  0.05    &     1.00 $\pm$  0.00    &    49.83 $\pm$  0.40      \\ 
   14    &    F606W    &    19.01 $\pm$  0.02    &   0.79 $\pm$  0.06    &   3.32 $\pm$  0.25    &   4.42 $\pm$  0.39    &     0.69 $\pm$  0.09    &     2.00 $\pm$  0.00    &    26.03 $\pm$  0.30       \\
         &    F814W    &    19.47 $\pm$  0.01    &   0.81 $\pm$  0.06    &   3.40 $\pm$  0.25    &   4.13 $\pm$  0.36    &     0.77 $\pm$  0.10    &     3.00 $\pm$  0.00    &   184.38 $\pm$  0.31      \\ 
   17    &    F606W    &    21.12 $\pm$  0.02    &   0.96 $\pm$  0.10    &   4.06 $\pm$  0.44    &   5.37 $\pm$  0.85    &     0.69 $\pm$  0.12    &     2.00 $\pm$  0.00    &    24.60 $\pm$  0.20       \\
         &    F814W    &    ...                  &  ...                  &  ...                  &  ...                  &    ...                  &    ...                  &    ...                    \\ 
   18    &    F606W    &    18.43 $\pm$  0.03    &   0.50 $\pm$  0.05    &   2.12 $\pm$  0.19    &   6.77 $\pm$  0.75    &     0.29 $\pm$  0.03    &     1.00 $\pm$  0.00    &    29.73 $\pm$  0.26       \\
         &    F814W    &    19.01 $\pm$  0.03    &   0.71 $\pm$  0.09    &   3.00 $\pm$  0.36    &  10.53 $\pm$  1.92    &     0.30 $\pm$  0.04    &     1.00 $\pm$  0.00    &   187.80 $\pm$  0.27      \\ 
   19    &    F606W    &    18.63 $\pm$  0.01    &   0.25 $\pm$  0.01    &   1.04 $\pm$  0.04    &   2.71 $\pm$  0.09    &     0.60 $\pm$  0.11    &     2.00 $\pm$  0.00    &    46.14 $\pm$  0.30       \\
         &    F814W    &    18.18 $\pm$  0.02    &   0.29 $\pm$  0.01    &   1.21 $\pm$  0.05    &   2.99 $\pm$  0.13    &     0.50 $\pm$  0.07    &     2.00 $\pm$  0.00    &    52.59 $\pm$  0.38      \\ 
   22    &    F606W    &    18.95 $\pm$  0.00    &   0.77 $\pm$  0.02    &   3.27 $\pm$  0.09    &   3.78 $\pm$  0.13    &     0.84 $\pm$  0.05    &     3.00 $\pm$  0.00    &    46.77 $\pm$  0.30       \\
         &    F814W    &    18.23 $\pm$  0.00    &   0.83 $\pm$  0.03    &   3.48 $\pm$  0.11    &   3.94 $\pm$  0.16    &     0.84 $\pm$  0.05    &     3.00 $\pm$  0.00    &    54.90 $\pm$  0.43      \\ 

\hline\hline
\multicolumn{10}{l}{Note. -- Table \ref{Table3} is published in its entirety in the electronic edition of the Journal.} \\ 
\multicolumn{10}{l}{$^{\mathsf{a}}$ Identification number in Paper I} \\
\end{longtable}
\normalsize
\renewcommand{\arraystretch}{1.0}
\end{landscape}




\end{document}